\documentclass[aps,prd,preprintnumbers,nofootinbib,onecolumn,a4paper]{revtex4-2}%
\usepackage[pdftex]{graphicx}
\usepackage{bm,latexsym,amsmath,amssymb,amsfonts,mathrsfs}
\usepackage{color}
\usepackage{subfigure}
\input{colordvi.tex}

\begin{document}

\title{\large 
Gravitational waves in Kasner spacetimes and Rindler wedges in Regge-Wheeler gauge: Formulation of Unruh effect}%
 
\author{Yuuki Sugiyama}
 \email{sugiyama.yuki@phys.kyushu-u.ac.jp}
\affiliation{Department of Physics, Graduate School of Science, Hiroshima University, Higashi-Hiroshima 739-8526, Japan}
\affiliation{
Department of Physics, 
Kyushu University, 744 Motooka, Nishi-Ku, Fukuoka 819-0395, Japan}
  
\author{Kazuhiro Yamamoto}
 \email{yamamoto@phys.kyushu-u.ac.jp}
\affiliation{Department of Physics, 
Kyushu University, 744 Motooka, Nishi-Ku, Fukuoka 819-0395, Japan}

\author{Tsutomu Kobayashi}
 \email{tsutomu@rikkyo.ac.jp}
\affiliation{Department of Physics, 
Rikkyo University, Toshima, Tokyo 171-8501, Japan}

\begin{abstract}
\centerline{\large Abstract}
We derive the solutions of gravitational waves in the future (F) expanding and past (P) shrinking Kasner spacetimes, as well as in the left (L) and right (R) Rindler wedges in the Regge-Wheeler gauge. The solutions for all metric components are obtained in an analytic form in each region. We identify the master variables, which are equivalent to massless scalar fields, to describe the gravitational degrees of freedom for the odd-parity and even-parity modes under the transformation in the two-dimensional plane-symmetric space. Then, the master variables are quantized, and we develop the quantum field theory of the gravitational waves in the F, P, L, and R regions. We demonstrate that the mode functions of the quantized gravitational waves in the left and right Rindler wedges are obtained by an analytic continuation of the left- moving and right-moving wave modes in Kasner spacetime. On the basis of these analyses, we discuss the Unruh effect of the quantized gravitational waves for an observer in a uniformly accelerated motion in Minkowski spacetime in an explicit manner for the first time. 
\end{abstract}

\preprint{RUP-20-35}

\maketitle

\section{Introduction}

Gravitational waves (GWs), predicted in general relativity, were directly detected by Advanced LIGO in 2015 for the first time. 
The first event, GW150914, is thought to come from a merger of binary black holes \cite{LIGO}. 
Gravitational waves are useful not only for testing general relativity itself, but also for exploring black hole physics. 
The detection of gravitational waves from a coalescence of a neutron star binary has enhanced their importance as astrophysical tools. 
Electromagnetic counterparts of gravitational-wave events such as neutron star merger events have been explored at various wavelengths, i.e., optical waves, infrared waves, x rays, and $\gamma$ rays (see 
e.g., \cite{GWNS}). The role of gravitational waves in cosmology is also important for exploring the primordial Universe. Primordial gravitational waves can be generated in an inflationary era from vacuum fluctuations, which might be detected in polarization anisotropies in the cosmic microwave background as a smoking gun of inflation, although they have not been detected \cite{Planck}.

The standard model of inflation yields a homogeneous and isotropic Universe satisfying the cosmological principle. Observations show that the Universe on very large scales appears to be consistent with the cosmological principle. However, some authors claim that the cosmic microwave background radiation shows anomalous features on large scales, such as hemispherical power asymmetry and dipole modulation \cite{cmb,fosalba}.
Models of inflation that may explain the large-scale anomalies have been proposed. For example, in Refs.~\cite{WKS,Soda} the authors proposed anisotropic inflation during which a $U(1)$ gauge field plays an important role in producing anisotropic expansion. They also investigated the cosmological perturbations in the anisotropic inflation models. Various anisotropic inflation models have been proposed so far, and cosmological perturbations have been found to be useful for characterizing the features of each anisotropic inflation model in terms of, e.g., the matter power spectrum and non-Gaussian features (see Refs.~\cite{DKP,Review}).

Many authors have investigated the behavior of gravitational waves in a universe that breaks spatial homogeneity or spatial isotropy \cite{TomitaDen,Pereira,Pitrou,Emir}. Cho and Speliotopoulos investigated the propagation of gravitational waves in a Bianchi type-I (B-I) universe \cite{Cho}. The authors of Refs. \cite{Sendouda1,Sendouda2} investigated the primordial gravitational waves from a preinflationary era in a B-I universe. Gravitational waves in a universe that breaks the cosmological principle are not trivial. For example, scalar, vector, and tensor modes do not decouple. The evolution of metric perturbations is not easy to solve in an analytic manner, and therefore numerical approaches or an approximation of small anisotropy are often taken.
In order to explore gravitational waves in an anisotropic universe or noninertial space, we focus on gravitational waves in the future expanding and past shrinking Kasner spacetimes as well as those in the left and right Rindler wedges. Kasner spacetime is a special case of a B-I universe, which is one of the simplest models of an anisotropic universe \cite{kasner}. The Rindler metric is a patch of Minkowski spacetime described by the coordinates of uniformly accelerating observers. Minkowski spacetime is covered by the future (F) expanding and past (P) shrinking Kasner spacetimes and the left (L) and right (R) Rindler wedges (see Fig.~\ref{fig:epsart})). One of the aims of the present paper is to investigate the quantum aspects of gravitational waves in Kasner spacetimes and Rindler wedges including the Unruh effect.

Hawking radiation is a well-known prediction of quantum field theory in a black hole spacetime \cite{hawking}. The Unruh effect predicts that a uniformly accelerating observer in Minkowski spacetime sees the vacuum state for an inertial observer as a thermally excited state with temperature $a/2\pi$, where a is the acceleration \cite{unruh}. The Unruh effect is an analogy of Hawking radiation through the equivalence principle, which is described on the basis of quantum field theory in Rindler wedges. The Unruh effect is explained by the fact that the Minkowski vacuum state is expressed as an entangled state between the left and right Rindler wedges when it is constructed on the Rindler vacuum. This is well known for the case of a scalar field and a Dirac field \cite{mode,Entangle,Ueda}. In the present paper, we investigate the vacuum structure of gravitational waves in Minkowski spacetime by finding the explicit expression of the solution of the tensor modes in the F, P, R, and L regions, which is a generalization of the works on a scalar field and a Dirac field to gravitational waves. Our result presents a formulation of the Unruh effect of gravitational waves in an explicit manner for the first time, as far as we know. This is achieved by extensively using the Regge-Wheeler gauge which is often used in a spherically symmetric spacetime \cite{RWgauge} for a plane-symmetric spacetime.

The present paper is organized as follows. In Sec. II, we first derive the action for the master variables of the gravitational waves in the F, P, R, and L regions using the Regge-Wheeler gauge for the odd modes and even modes, respectively \cite{RWgauge}. The resultant action for the master variables is found to be equivalent to that of a massless scalar field, which is quantized in Sec. III. Then, we discuss the Bogoliubov transformation for two different sets of the solutions, and we show that the solutions in Kasner space- times (F and P regions) are analytically continued to the left and right Rindler wedges (L and R regions). In Sec. IV, we demonstrate the analytic continuations of the metric components of the F and P regions to those of the L and R regions. Then, we demonstrate that the Minkowski vacuum state is described as an entangled state between the two modes described in the whole region of the Minkowski spacetime, which leads to a description of the Unruh effect of the gravitational waves. In Sec. V, we demonstrate the calculation of the expectation values of the energy density of the vacuum fluctuations of the gravitational waves associated with the Rindler vacuum state. Section \ref{secVI} is devoted to summary and conclusions. In Appendix \ref{Minv}, we briefly review the positive-mode frequency function of the Minkowski vacuum state in the F region. In Appendix \ref{secv}, we summarize the results of Sec. IV, i.e., the analytic continuation property of the metric perturbation of the gravitational waves in the F, P, R, and L regions in the Regge-Wheeler gauge. In Appendix \ref{eneden}, we present some details of the calculation in Sec. \ref{secV}. Throughout the present paper,we use the natural units $\ c=\hbar=k_B=1$.

\def\vec{\bm}
%%%%%%%%%%%%%%%%%%%%%%%%%%%%%%%%%%%%%%%%%%%
\section{Classical solutions of GWs in Rindler and Kasner spacetime}
%%%%%%%%%%%%%%%%%%%%%%%%%%%%%%%%%%%%%%%%%%%

\subsection{GWs in future expanding Kasner spacetime (F region)}

\begin{figure}[t]
\includegraphics[width=0.5\textwidth=3cm]{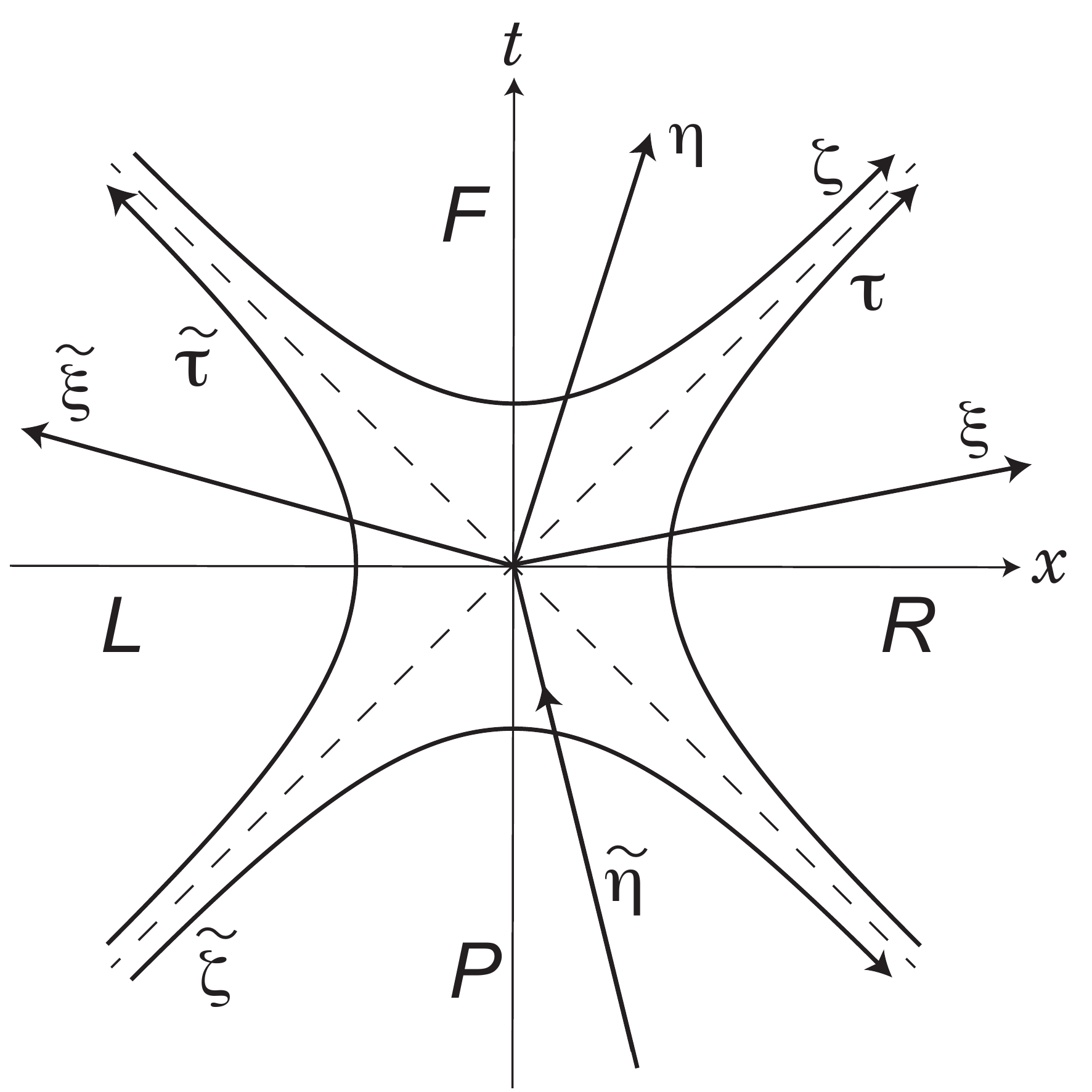}% Here is how to import EPS art
\caption{\label{fig:epsart} Four regions of Minkowski spacetime and corresponding coordinates.}
\end{figure}

We start with deriving the solution of GWs in the future expanding Kasner spacetime
(the F region) by using the Regge-Wheeler gauge.
The line element of the F region is given by
\begin{align}
ds^2=e^{2a\eta}(-d\eta^2+d\zeta^2)+dy^2+dz^2\equiv \bar g_{\mu\nu}^Fdx^\mu dx^\nu,
\label{kasner-F-met}
\end{align}
where $-\infty<\eta<\infty$, $-\infty<\zeta<\infty$, and 
$a$ is a constant. The Kasner coordinates of the F region
are related to the global coordinates of Minkowski spacetime as
\begin{align}
t=\frac{1}{a}e^{a\eta}\cosh{a\zeta},\quad x=\frac{1}{a}e^{a\eta}\sinh{a\zeta},
\end{align}
as shown in Fig.~\ref{fig:epsart}. 
The metric perturbations $h_{{\mu}{\nu}}^F$ are defined by
\begin{align}
       g_{{\mu}{\nu}}^F = \bar{g}_{{\mu}{\nu}}^F + h_{{\mu}{\nu}}^F.\  
\end{align}

Following \cite{RWgauge}, we write
\begin{align}
  h^{F}_{0a} &= \partial_a{h^{F}_0} + \epsilon_{ab} \partial^b{\chi^{F}},\\
  h^{F}_{1a} &= \partial_a{h^{F}_1} + \epsilon_{ab} \partial^b{\psi^{F}},\\
  h^{F}_{ab} &= h^{F}\delta_{ab}+ \partial_a {\partial_b {h^{F}_3}} + \epsilon_{c(a}\partial_{b)}\partial^c{\gamma^{F}} ,
\end{align}
where $a$, $b$, and $c$ %run from $2$ to $3$, then $x^a$ and $x^b$ denote $y$ or $z$.
stand for 2 or 3, and $\epsilon_{ab}$ is the completely antisymmetric tensor: $\epsilon_{22}=\epsilon_{33}=0,\ \epsilon_{23}=1=-\epsilon_{32}$.
Therefore, the metric perturbations are written using
the $10$ functions,  $(h^{F}_{00},h^{F}_{01},h^{F}_{11},h^{F}_0,h^{F}_1,\chi^{F},\psi^{F},h^{F},h^{F}_3,\gamma^{F})$. 
With respect to the parity transformation on the $y$-$z$ plane, 
these metric perturbations can be classified into the 
odd modes described by $(\chi^{F},\psi^{F},\gamma^{F})$ and 
the even modes described by $(h^{F}_{00},h^{F}_{01},h^{F}_{11},h^{F}_0,h^{F}_1,h^{F},h^{F}_3)$.
%%%%%%%%%%%%%%%%
\label{secII}
%%%%%%%%%%%

Let us consider the gauge transformation
generated by the infinitesimal transformation generator 
\begin{align}
K_{\mu}=K^{\rm odd}_{\mu}+K^{\rm even}_{\mu}
\end{align}
with
\begin{align}
K^{\rm odd}_{\mu} \equiv 
\left(
\begin{array}{c}
0\\
0\\
\epsilon_{ab}\partial^b{\lambda}
\end{array}
\right),\quad
K^{\rm even}_{\mu} \equiv
\left(
\begin{array}{c}
\delta \eta \\
\delta \zeta \\
\partial_a{\delta x}
\end{array}
\right),
\end{align}
where $K^{\rm odd}_{\mu}$ and $K^{\rm even}_{\mu}$ are the 
generator of the gauge transformation for the odd
and even modes, respectively. 
Using this gauge freedom,
we take the Regge-Wheeler gauge in which
$h^{F}_0=h^{F}_1=h^{F}_3=\gamma^{F}=0$. 
The metric perturbations are then described by
the $6$ functions $(h^{F}_{00},h^{F}_{01},h^{F}_{11},h^{F},\chi^{F},\psi^{F})$:
\begin{align}
\label{TTb}
{h}^{F}_{\mu\nu} = 
\begin{pmatrix}
h^{F}_{00}&h^{F}_{01}&\partial_z{\chi^{F}}&-\partial_y{\chi^{F}}\\
h^{F}_{01}&h^{F}_{11}&\partial_z{\psi^{F}}&-\partial_y{\psi^{F}}\\
\partial_z{\chi^{F}}&\partial_z{\psi^{F}}&h^{F}&0\\
-\partial_y{\chi^{F}}&-\partial_y{\psi^{F}}&0&h^{F}
\end{pmatrix}.
\end{align}
%In this appendix, we

Let us now derive the quadratic Lagrangians
for the master variables
for odd and even parity perturbations.
Although we focus on the F region,
the derivation can be extended straightforwardly to
the P, R, and L regions.

\subsubsection{Odd parity perturbations}\label{odd}
%To simplify the notation we omit the label $F$ from the variables.
Around the future expanding Kasner metric
we expand the Einstein-Hilbert Lagrangian,
$\sqrt{-g}R/2$, to second order in metric perturbations.
Performing integration by parts, we obtain
\begin{align}
8\pi G \mathscr{L}^{\rm odd} 
= \frac{1}{4}e^{-2a\eta} \left[
(\partial_{\perp}{\partial_{\eta}{\psi}^{F}})^2-2(\partial_{\perp}\partial_{\zeta}{\chi}^{F})(\partial_{\perp}{\partial_{\eta}{\psi}^{F}}) + (\partial_{\perp}\partial_{\zeta}{\chi}^{F})^2
\right]
{}-\frac{1}{4}(\partial^2_{\perp}{\psi}^{F})^2 + \frac{1}{4}(\partial^2_{\perp}{\chi}^{F})^2, 
\label{L2-1}
\end{align}
%where a dot and a prime denote differentiation with respect to $\eta$ and $\xi$, respectively.
By introducing the auxiliary field $\phi^{F}$,
the above Lagrangian can be written equivalently as
\begin{align}
 8\pi G \mathscr{L}^{\rm odd}= \frac{1}{4}\left[
 -e^{2a\eta}(\partial_{\perp}{\phi}^{F})^2 + 2\partial_{\perp}{\phi}^{F}(\partial_{\perp}{\partial_{\eta}{\psi}^{F}} - \partial_{\perp}{\partial_{\zeta}\chi}^{F})\right] -\frac{1}{4}(\partial^2_{\perp}{\psi}^{F})^2 + \frac{1}{4}(\partial^2_{\perp}{\chi}^{F})^2.
 \label{L2-2}
\end{align}
Indeed, from the Euler-Lagrange equation for $\phi^{F}$ one has
\begin{align}
\phi = e^{-2a\eta}(\partial_{\eta}{\psi}^{F} - \partial_{\zeta}\chi^{F}),
\end{align}
using which one can remove $\phi^{F}$ from Eq.~\eqref{L2-2}
to reproduce the original Lagrangian~\eqref{L2-1}.

Now, the Euler-Lagrange equations for $\psi^{F}$ and $\chi^{F}$
derived from Eq.~\eqref{L2-2} read, respectively,
\begin{align}
\partial^2_{\perp}\psi^{F}= \partial_{\eta}{\phi}^{F},\quad
\partial^2_{\perp}\chi^{F} = \partial_{\zeta}\phi^{F}.\label{constraintsodd}
\end{align}
Substituting these constraint equations back to Eq.~\eqref{L2-2},
we arrive at the quadratic Lagrangian for the dynamical variable $\phi^{F}$,
\begin{align}
8\pi G \mathscr{L}^{\rm odd}=\frac{1}{4}\left[({\partial_{\eta}{\phi}^{F}})^2 - ({\partial_{\zeta}\phi}^{F})^2 - e^{2a\eta}(\partial_{\perp}{\phi}^{F})^2\right],
\label{Loddfinal}
\end{align}
where we defined $(\partial_\perp \phi^{F})^2=(\partial_y\phi^{F})^2+(\partial_z\phi^{F})^2$ (and similarly for $(\partial_\perp h^{F})^2$ below).
Thus, the odd parity sector is found to be governed by the single master variable $\phi^{F}$.
The metric perturbations in the Regge-Wheeler gauge, $\psi^{F}$ and $\chi^{F}$,
are determined in terms of $\phi^{F}$ through the constraint equations~\eqref{constraintsodd}.

\subsubsection{Even parity perturbations}\label{even}
We also expand the Einstein-Hilbert Lagrangian to second order in even parity perturbations to get
\begin{align}
8\pi G\mathscr{L}^{\rm even}
 &= - \frac{1}{4}({\partial_{\eta}{h}^{F}})^2 + \frac{1}{2}e^{-2a\eta}{h_{11}^{F}}\partial^2_{\eta}{h}^{F} +ae^{-2a\eta}{h_{01}^{F}}\partial_{\zeta}{h}^{F} -e^{-2a\eta}{h_{01}^{F}}\partial_{\eta}\partial_{\zeta}{h}^{F} + \frac{1}{4}e^{-2a\eta}(\partial_{\perp}{h_{01}^{F}})^2 \nonumber\\
&{}+ \frac{1}{4}\partial_{\perp}{h_{11}^{F}} \partial_{\perp}{h}^{F} + \frac{1}{4}({\partial_{\zeta}h}^{F})^2  -\frac{a}{2}e^{-2a\eta}h_{11}^{F}\partial_{\eta}{h}^{F}\nonumber\\
&{} + h_{00}^{F}\left(\frac{1}{4}e^{-2a\eta}{\partial^2_{\perp}{h_{11}^{F}}} + \frac{1}{4}\partial^2_{\perp}{h}^{F} + \frac{1}{2}e^{-2a\eta}\partial^2_{\zeta}{h}^{F} - \frac{a}{2}e^{-2a\eta}\partial_{\eta}{h}^{F}\right),
 \end{align}
 where we performed integration by parts and omitted
total derivatives.
From the Euler-Lagrange equations for $h_{00}^{F}$, $h_{01}^{F}$, and $h_{11}^{F}$,
we obtain the following constraint equations:
\begin{align}
    \partial^2_{\perp}{h_{11}^{F}} &= -e^{2a\eta}\partial^2_{\perp}{h}^{F} + 2a{\partial_{\eta}{h}^{F}} - 2{\partial^2_{\zeta}h^{F}},\label{oddconst1}
    \\
    \partial^2_{\perp}{h_{01}^{F}} &=  -2(\partial_{\eta}\partial_{\zeta}{h^{F}} - a\partial_{\zeta}h^{F}),
    \\
    \partial^2_{\perp}{h_{00}^{F}} &= e^{2a\eta}\partial^2_{\perp}{h}^{F} + 2a\partial_{\eta}{h}^{F}- 2\partial^2_{\eta}{h}^{F}.
    \label{oddconst3}
\end{align}
These equations can be used to eliminate $h_{11}^{F}$, $h_{01}^{F}$, and $h_{11}^{F}$
from the Lagrangian. After integration by parts, we end up with
the quadratic Lagrangian for the single master variable $h^{F}$,
  \begin{align}
  8\pi G \mathscr{L}^{\rm even}
  =  \frac{1}{4} \left[{(\partial_{\eta}{h}^{F})}^2 - (\partial_{\zeta}{h}^{F})^2 -e^{2a\eta}(\partial_{\perp} h^{F})^2\right],
  \end{align}
  which has essentially the same form as Eq.~\eqref{Loddfinal}.
  The other components of the even parity metric perturbations
  are determined through the constraint equations~\eqref{oddconst1}--\eqref{oddconst3}.

Thus, the quadratic action for the gravitational waves is
obtained from the Einstein-Hilbert action and can be written
in terms of the two decoupled master variables as
$S^{(2)}=S^F_{g(o)}+S^F_{g(e)}$,
\begin{align}
S^{F}_{g(\rm o)} &= \frac{1}{32\pi G}\int d\eta  d\zeta  dy  dz
\left[(\partial_{\eta}\phi^{F})^2 -(\partial_{\zeta}\phi^{F})^2 - e^{2a\eta}(\partial_\perp \phi^{F})^2\right],
\label{actionoddF}\\
S^{F}_{g(\rm e)} &= \frac{1}{32\pi G}\int d\eta d\zeta dy dz\left[(\partial_{\eta}h^{F})^2 -(\partial_{\zeta}h^{F})^2 - e^{2a\eta}(\partial_\perp h^{F})^2\right],
\label{actionevenF}
\end{align}
where $S^{F}_{g(o)}$ and $S^{F}_{g(e)} $ are the 
actions for the odd and even modes, respectively.
%and we defined $(\partial_\perp \phi^{F})^2=(\partial_y\phi^{F})^2+(\partial_z\phi^{F})^2$
%(and similarly for $(\partial_\perp h^{F})^2$).
Here $\phi^F$ itself is not a metric perturbation variable,
but it is related to $\psi^F$ and $\chi^F$ in a nontrivial way.
One may notice that $\phi^F$ and $h^F$ are subject to
the same action as that of a massless scalar field
living in the Kasner spacetime~\eqref{kasner-F-met}.
Once the solutions for $\phi^F$ and $h^F$ are obtained,
all the other variables are determined in terms of these two
master variables through the constraint equations, as demonstrated below.
This is as it should be, because we have only two dynamical degrees of
freedom in general relativity.
%These actions are obtained by substituting the metric perturbations to the Einstein-Hilbert action, expanding it to quadratic order, and performing integration of parts.
%The detailed derivation is presented in Appendix A.

Let us then proceed to solving the equations of motion.
We obtain the following equation of motion for the odd mode,
\begin{align}
\label{eom1}
\left[-{\partial^2_\eta}+{ \partial_\zeta^2}
+e^{2a\eta}{ \partial_\perp^2}\right]\phi^{F}=0,
\end{align}
where 
$\partial_\perp^2=\partial_y^2+\partial_z^2$.
The odd-mode metric perturbations, $\psi^F$ and $\chi^F$, are
given in terms of $\phi^F$ through the constraint equations \eqref{constraintsodd}.
%\begin{align}
%\partial^2_\perp{\psi}^{F} = \partial_{\eta}\phi^{F} ,\quad
%\partial^2_\perp{\chi}^{F} = \partial_{\zeta}\phi^{F}.
%\label{oddmetic1}
%\end{align}
%(See Appendix~\ref{odd} for the derivation.)
When $\phi^{F}$ is written in a mode-expanded form
\begin{align}
   \phi^{F}=\varphi^{F,{\rm o}}(\eta) e^{i (k_x\zeta+\bm k_\perp\cdot \bm x_\perp) },
\end{align}
the coefficient $\varphi^{F,{\rm o}}(\eta)$ obeys
\begin{align}
[\partial_\eta^2+k^2(\eta)]\varphi^{F,{\rm o}}(\eta) =0,
\label{equationvarFo}
\end{align}
where we defined $k^2(\eta)=k_x^2+e^{2a\eta}\kappa^2$ with $\kappa^2=|{\bm k}_\perp|^2$. 
Then, using Eq.~(\ref{constraintsodd}), we have
\begin{align}
\psi^{F} &=-\frac{\partial_\eta \varphi^{F,{\rm o}}}{\kappa^2} e^{i (k_x\zeta+\bm k_\perp\cdot \bm x_\perp) }, 
\\
\chi^F &=-\frac{i k_x\varphi^{F,{\rm o}}}{\kappa^2} e^{i (k_x\zeta+\bm k_\perp\cdot \bm x_\perp) }.
\end{align}

For the even mode, $h^F$, we have the same equation of motion as Eq. (\ref{eom1}).
The other even-mode metric perturbations are given in terms of $h^{F}$ through 
\begin{align}
\partial^2_\perp{{h}^{F}_{00}} &= e^{2a\eta}\partial^2_\perp{{h}}^{F} + 2a{\partial_{\eta}{{h}}}^{F} - 2\partial^2_{\eta}{{h}}^{F},
\label{oddmetrice1}\\
\partial^2_\perp{{h}^{F}_{01}} &=  -2\partial_{\zeta}(\partial_{\eta}{{h}}^{F} - a{h}^{F}),
\label{oddmetrice2}
\\
\partial^2_\perp{{h}^{F}_{11}} &= -e^{2a\eta}\partial^2_\perp{{h}}^{F} + 2a\partial_{\eta}{{h}}^{F} - 2\partial^2_{\zeta}{{h}}^{F}.
\label{oddmetrice3}
\end{align}
%(See Appendix~\ref{even} for the derivation.)
Then, we write $h^F$ in a mode-expanded form
\begin{align}
   h^{F}=\varphi^{F,{\rm e}}(\eta) e^{i (k_x\zeta+\bm k_\perp\cdot \bm x_\perp) }, 
\end{align}
and $\varphi^{F,{\rm e}}(\eta)$ obeys the same equation as
 Eq.~(\ref{equationvarFo}),
\begin{align}
[\partial_\eta^2+k^2(\eta)]\varphi^{F,{\rm e}}(\eta) =0.
\label{equationvarFe}
\end{align}
From Eqs. (\ref{oddmetrice1}), (\ref{oddmetrice2}), and (\ref{oddmetrice3}), we have
\begin{align}
\label{hmunuF}
{h}^{F}_{00}&=\left(e^{2a\eta}\varphi_{}^{F,\rm e}
-\frac{2a}{\kappa^2}\partial_{\eta}{\varphi}_{}^{F,\rm e}
+\frac{2}{\kappa^2}\partial^2_{\eta}{\varphi}_{}^{F,\rm e}\right)e^{i (k_x\zeta+\bm k_\perp\cdot \bm x_\perp) },
\\
{h}^{F}_{01}&=
\frac{2 i k_x}{\kappa^2}(\partial_{\eta}{\varphi}_{}^{F,\rm e}
-a\varphi_{}^{F,\rm e})e^{i (k_x\zeta+\bm k_\perp\cdot \bm x_\perp) },
\\
{h}^{F}_{11}&=\left(-\frac{k^2(\eta)}{\kappa^2}\varphi_{}^{F,\rm e}
-\frac{k_x^2}{\kappa^2}\varphi_{}^{F,\rm e}
-\frac{2a}{\kappa^2}\partial_{\eta}{\varphi}_{}^{F,\rm e}\right)e^{i (k_x\zeta+\bm k_\perp\cdot \bm x_\perp) }.
\end{align}

Finally, summarizing the above results, the metric perturbations
in the F region in the Regge-Wheeler gauge are given by
\begin{align}
h_{\mu\nu}^F=({\widetilde h}_{\mu\nu}^{F,\rm o}+{\widetilde h}_{\mu\nu}^{F,\rm e})e^{i (k_x\zeta+\bm k_\perp\cdot \bm x_\perp) }
\end{align}
with
\begin{align}
\widetilde h_{\mu\nu}^{F,\rm o}&=
\left(
\begin{array}{cccc}
0 &0 &\displaystyle{ \frac{k_x k_z}{\kappa^2}\varphi_{}^{F,\rm o}} &\displaystyle{ -\frac{k_x k_y}{\kappa^2}\varphi_{}^{F,\rm o}}\\
* &0& \displaystyle{-\frac{i k_z}{\kappa^2}\partial_{\eta}\varphi_{}^{F,\rm o}}&\displaystyle{ \frac{i k_y}{\kappa^2}\partial_{\eta}\varphi_{}^{F,\rm o} }\\
*&* &0 &0\\
* &* & *& 0
\end{array}\right) ,
\label{hmunuFo}
\\
\widetilde h_{\mu\nu}^{F,\rm e}&=
\left(
\begin{array}{cccc}
\displaystyle{
e^{2a\eta}\varphi^{F,\rm e}-\frac{2a}{\kappa^2}\partial_{\eta}\varphi_{}^{F,\rm e}+\frac{2}{\kappa^2}\partial^2_{\eta}\varphi_{}^{F,\rm e}} & \displaystyle{\frac{2 i k_x}{\kappa^2}(\partial_{\eta}\varphi_{}^{F,\rm e}-a\varphi_{}^{F,\rm e})} &0 &0\\
* &\displaystyle{-\frac{k^2(\eta)}{\kappa^2}\varphi_{}^{F,\rm e}
-\frac{k_x^2}{\kappa^2}\varphi_{}^{F,\rm e} -\frac{2a}{\kappa^2}\partial_{\eta}\varphi_{}^{F,\rm e} }& 0&0\\
*&* &\varphi_{}^{F,\rm e} &0\\
* &* & *& \varphi_{}^{F,\rm e}
\end{array}\right).
\label{hmunuFe}
\end{align}
We can thus determine all the metric components simply by
solving a massless Klein-Gordon equation in the Kasner spacetime.

\begin{table}[b]
\label{connect}
\begin{tabular}{lll}
\hline 
\hline\\
\vspace{2mm}
$\mathrm{F} \to \mathrm{R}\quad $ & $\displaystyle{ \tau=\zeta-\frac{\pi}{2 a} i}$, \quad & $ \displaystyle{ \xi=\eta+\frac{\pi}{2 a} i } $\\ 
\vspace{2mm}
$\mathrm{F} \to \mathrm{L}$ & $\displaystyle{ \tilde{\tau}=-\zeta-\frac{\pi}{2 a} i,}$ \quad &  $\displaystyle{\tilde{\xi}=\eta+\frac{\pi}{2 a} i} $\\ 
\vspace{2mm}
$\mathrm{P} \to \mathrm{R}$ & $\displaystyle{ \tau=-\tilde{\zeta}-\frac{\pi}{2 a} i}$,  \quad & $\displaystyle{\xi=-\tilde{\eta}-\frac{\pi}{2 a} i}$ \\
\vspace{2mm}
$\mathrm{P} \to \mathrm{L}$ & $\displaystyle{ \tilde{\tau}=\tilde{\zeta}-\frac{\pi}{2 a} i}$, \quad &$\displaystyle{\tilde{\xi}=-\tilde{\eta}-\frac{\pi}{2 a} i }$\\ 
\hline
\end{tabular}
\caption{Analytic continuations and the relations among the 
coordinates in the F, P, R, and L regions.
A more detailed discussion is given in Sec.~\ref{cal}.}
\end{table}

\subsection{GWs in past shri{n}king Kasner spacetime (P region)}

Let us then derive the solution of GWs in the past shri{n}king Kasner spacetime (the
P region) in the Regge-Wheeler gauge.
The analysis presented in this and subsequent subsections
follows closely, and hence has some overlaps with the one in the previous subsection.
The line element of the P-region is written as
\begin{align}
ds^2=e^{-2a\widetilde{\eta}}(-d\widetilde{\eta}^2+d\widetilde{\zeta}^2)+dy^2+dz^2\equiv \bar g_{\mu\nu}^P dx^{\mu}dx^{\nu},
\end{align}
where the coordinates in the P region is related to those in 
the global Minkowski coordinates as 
\begin{align}
t=-\frac{1}{a}e^{-a\widetilde{\eta}}\cosh{a\widetilde{\zeta}},\quad  x=\frac{1}{a}e^{-a\widetilde{\eta}}\sinh{a\widetilde{\zeta}},
\end{align}
as shown in Fig.~\ref{fig:epsart}.
The coordinates in the P region is obtained by 
the analytic continuation of the coordinates in the F region as $\zeta=-\widetilde\zeta$ and $\eta=-\widetilde\eta-\pi i/a$ (TABLE I). 
We can write the metric perturbations in the P region in the Regge-Wheeler gauge as
$
       g_{{\mu}{\nu}}^P = \bar{g}_{{\mu}{\nu}}^P + h_{{\mu}{\nu}}^P
$
with
\begin{align}
\label{TTc}
{h}^{P}_{\mu\nu} = 
\begin{pmatrix}
{h}^{P}_{00}&{h}^{P}_{01}&\partial_z{\chi}^{P}&-\partial_y{\chi}^{P}\\
{h}^{P}_{01}&{h}^{P}_{11}&\partial_z{\psi}^{P}&-\partial_y{\psi}^{P}\\
\partial_z{\chi}^{P}&\partial_z{\psi}^{P}&{h}^{P}&0\\
-\partial_y{\chi}^{P}&-\partial_y{\psi}^{P}&0&{h}^{P}
\end{pmatrix}.
\end{align}

The master variable for the odd modes, $\phi^P$,
can be found in a similar way to those in the F region, giving the action and equation of motion
\begin{align}
S^{P}_{g(\rm o)} &= \frac{1}{32\pi G}\int d\tilde{\eta} d\tilde{\zeta}  dy  dz\left[(\partial_{\tilde{\eta}}\phi^{P})^2 -(\partial_{\tilde{\zeta}}\phi^{P})^2 - e^{-2a\tilde{\eta}}(\partial_\perp \phi^{P})^2\right],
\label{actionSp}
\end{align}
\begin{align}
\left[-{ \partial_{\tilde{\eta}}^2}+{ \partial_{\tilde{\zeta}}^2}
+e^{-2a\tilde{\eta}}{ \partial_\perp^2}\right]\phi^P=0.
\label{oddequationV}
\end{align}
The master variable of the even mode, $h^P$, has the same action and
hence obeys the same
equation of motion as Eqs. (\ref{actionSp})
and (\ref{oddequationV}). The metric perturbations are given by solving 
\begin{align}
\partial^2_\perp{{\psi}}^{P} = \partial_{\tilde{\eta}}{{{\phi}}}^{P},\quad
\partial^2_\perp{{\chi}}^{P} = \partial_{\tilde{\zeta}}{{{\phi}}}^{P},
\label{evenmetric1}
\end{align}
and 
\begin{align}
\partial_\perp^2{{h}^{P}_{00}} &= e^{-2a\tilde{\eta}}\partial_\perp^2{{h}}^{P} - 2a{\partial_{\tilde{\eta}}{{h}}}^{P} - 2\partial^2_{\tilde{\eta}}{{h}}^{P},\\
\partial_\perp^2{{h}^{P}_{01}} &=  -2\partial_{\tilde{\zeta}}(\partial_{\tilde{\eta}}{{h}}^{P} + a{h}^{P}),\\
\partial_\perp^2{{h}^{P}_{11}} &= -e^{-2a\tilde{\eta}}\partial_\perp^2{{h}}^{P} - 2a\partial_{\eta}{{h}}^{P} - 2\partial^2_{\tilde{\zeta}}{{h}}^{P}.
\label{evenmetric2}
\end{align}

When $\phi^P$ and $h^P$ are written in the forms
\begin{align}
\phi^P=\varphi^{P,\rm o}(\widetilde\eta)e^{i(k_x\widetilde\zeta+\bm k_\perp\cdot\bm x_\perp)}, 
\quad 
h^P=\varphi^{P,\rm e}(\widetilde\eta)e^{i(k_x\widetilde\zeta+\bm k_\perp\cdot\bm x_\perp)}, 
\end{align}
we have the metric perturbations in the P~region in the Regge-Wheeler 
gauge,
\begin{align}
h_{\mu\nu}^P=(\widetilde h_{\mu\nu}^{P,\rm o}+\widetilde h_{\mu\nu}^{P,\rm e})
e^{i(k_x\widetilde\zeta+\bm k_\perp\cdot\bm x_\perp)},
\end{align}
with
\begin{align}
\label{hmunuPo}
\widetilde h_{\mu\nu}^{P,\rm o}&=
\left(
\begin{array}{cccc}
0&0&\displaystyle{ \frac{k_x k_z}{\kappa^2}\varphi_{}^{P,\rm o}} &\displaystyle{ -\frac{k_x k_y}{\kappa^2}\varphi_{}^{P,\rm o}}\\
* &0& -\displaystyle{\frac{i k_z}{\kappa^2}\partial_{\tilde{\eta}}\varphi_{}^{P,\rm o}}&\displaystyle{ \frac{i k_y}{\kappa^2}\partial_{\tilde{\eta}}\varphi_{}^{P,\rm o} }\\
*&* &0 &0\\
* &* & *& 0
\end{array}
\right),
\\
\label{hmunuPe}
\widetilde h_{\mu\nu}^{P,\rm e}&=
\left(
\begin{array}{cccc}
\displaystyle{
e^{-2a\tilde\eta}\varphi_{}^{P,\rm e}+\frac{2a}{\kappa^2}\partial_{\tilde{\eta}}\varphi_{}^{P,\rm e}+\frac{2}{\kappa^2}\partial^2_{\tilde{\eta}}\varphi_{}^{P,\rm e}} & \displaystyle{\frac{2 i k_x}{\kappa^2}(\partial_{\tilde{\eta}}\varphi_{}^{P,\rm e}+a\varphi_{}^{P,\rm e})} &0 &0\\
* &\displaystyle{-\frac{\tilde{k}^2(\widetilde{\eta})}{\kappa^2}\varphi_{}^{P,\rm e}
-\frac{k_x^2}{\kappa^2}\varphi_{}^{P,\rm e} +\frac{2a}{\kappa^2}\partial_{\widetilde{\eta}}\varphi_{}^{P,\rm e} }& 0&0\\
*&* &\varphi_{}^{P,\rm e} &0\\
* &* & *& \varphi_{}^{P,\rm e}
\end{array}
\right),
\end{align}
where $\widetilde k^2(\widetilde\eta)=(k_x^2+\kappa^2e^{-2a\widetilde{\eta}})$. 

%%%%%%%%%%%%%%%%%%%%%%%%%%%%%%%%%%%%%%%%%%%%%%%%%%%%%
\subsection{GWs in right Rindler wedge (R region)}
%%%%%%%%%%%%%%%%%%%%%%%%%%%%%%%%%%%%%%%%%%%%%%%%%%%%%
\def\calk{{\cal K}}
Next, we derive the solution of GWs in the right Rindler wedge (the R region).
The line element in the R region is given by
\begin{align}
\label{metricR}
ds^2=e^{2a\xi}(-d\tau^2+d\xi^2)+dy^2+dz^2 \equiv \bar{g}^{R}_{\mu \nu}dx^{\mu}dx^{\nu},
\end{align}
where $a$ is a constant which is interpreted as
a uniform acceleration. The coordinates in the R~region is related to the coordinates of Minkowski spacetime as
\begin{align}
\label{coordR}
t=\frac{1}{a}e^{a\xi}\sinh{a\tau},\quad
x=\frac{1}{a}e^{a\xi}\cosh{a\tau},
\end{align}
as shown in Fig.~\ref{fig:epsart}.
The metric perturbations in the R~region can be written as
\begin{align}
g^{R}_{\mu \nu}=\bar{g}^{R}_{\mu \nu}+h^{R}_{\mu \nu}
\end{align}
with
\begin{align}
h^{R}_{\mu \nu} = 
\begin{pmatrix}
h^{R}_{00}&h^{R}_{01}&\partial_z{\chi}^{R}&-\partial_y{\chi}^{R}\\
h^{R}_{01}&h^{R}_{11}&\partial_z{\psi}^{R}&-\partial_y{\psi}^{R}\\
\partial_z{\chi}^{R}&\partial_z{\psi}^{R}&h^{R}&0\\
-\partial_y{\chi}^{R}&-\partial_y{\psi}^{R}&0&h^{R}
\end{pmatrix}.
\end{align}
The action for the odd parity master variable in the R region is given by
\begin{align}
S^{R}_{g(\rm o)} =\frac{1}{32\pi G}\int d{\tau} d{\xi} dy  dz\left[(\partial_{\tau}\phi^{R})^2 -(\partial_{\xi}\phi^{R})^2 - e^{2a{\xi}}(\partial_\perp \phi^{R})^2\right],
\label{actionRo}
\end{align}
which leads to the equation of motion
\begin{align}
\label{EOMR}
\left[-{ \partial_\tau^2}+{ \partial_\xi^2}
+e^{2a\xi}{ \partial_\perp^2}\right]\phi^R=0.
\end{align}
The metric perturbations $\psi^{R}$ and $\chi^{R}$ are related to $\phi^{R}$ as
\begin{align}
\partial_\perp^2\psi^{R} = \partial_{\tau}{\phi}^{R},\quad
\partial_\perp^2\chi^{R} = \partial_{\xi}\phi^{R}.
\end{align}
Writing $\phi^R$ in the form,
\begin{align}
\phi^{R}=\varphi^{R,\rm o}(\xi)e^{-i\omega\tau+i\bm k_\perp\cdot\bm x_\perp}, 
\end{align}
one sees that
$\varphi^{R,\rm o}$ obeys
\begin{align}
[\partial_\xi^2+{\cal K}^2(\xi) ]\varphi^{R,\rm o}=0
\end{align}
with ${\cal K}^2(\xi)=\omega^2-\kappa^2 e^{2a\xi}$.
Thus, also in the right Rindler wedge the problem reduces to
analyzing a system of a massless scalar field.

For the even mode, we obtain the same action as that of the odd mode,
\begin{align}
S^{R}_{g(\rm e)} =\frac{1}{32\pi G}\int d{\tau} d{\xi} dy dz\left[(\partial_{\tau}h^{R})^2 -(\partial_{\xi}h^{R})^2 - e^{2a{\xi}}(\partial_\perp h^{R})^2\right],
\label{actionRe}
\end{align}
and hence the equation of motion is also the same as that of $\phi^{R}$.
Using $h^R$ one can write the other metric functions as
\begin{align}
\partial_\perp^2{h^{R}_{00}} &= e^{2a\xi}\partial_\perp^2{h}^{R} + 2a\partial_{\xi}{h^{R}} - 2\partial^2_{\tau}{h}^{R},
\label{equationR1}
\\
\partial_\perp^2{h^{R}_{01}} &=  -2\partial_{\tau}(\partial_{\xi}{h^{R}} - a{h}^{R}),
\label{equationR2}
\\
\partial_\perp^2{h^{R}_{11}} &= -e^{2a\xi}\partial_\perp^2{h}^{R} + 2a\partial_{\xi}{h^{R}} - 2\partial^2_{\xi}{h^{R}}
\label{equationR3}.
\end{align}
Writing $h^R$ in the form,
\begin{align}
h^{R}=\varphi^{R,\rm e}(\xi)e^{-i\omega\tau+i\bm k_\perp\cdot\bm x_\perp}, 
\end{align}
the coefficient $\varphi^{R,\rm e}$ obeys
\begin{align}
[\partial_\xi^2+{\cal K}^2(\xi) ]\varphi^{R,\rm e}=0,
\end{align}
where ${\cal K}$ was already defined above.
The solutions of Eqs. (\ref{equationR1}), (\ref{equationR2}), (\ref{equationR3}) are
obtained as
\begin{align}
{h^{R}_{00}} &= \left(e^{2a\xi}{\varphi}^{R,\rm e} -\frac{2a}{\kappa^2}\partial_{\xi}{\varphi^{R,\rm e}} - 2\frac{\omega^2}{\kappa^2}{\varphi}^{R,\rm e}\right)e^{-i\omega\tau+i\bm k_\perp\cdot\bm x_\perp},\\
{h^{R}_{01}} &=  -\frac{2 i \omega}{\kappa^2}(\partial_{\xi}{\varphi^{R,\rm e}} - a{\varphi}^{R,\rm e})e^{-i\omega\tau+i\bm k_\perp\cdot\bm x_\perp},
\\
{h^{R}_{11}} &= \left(-e^{2a\xi}{\varphi}^{R,\rm e} - \frac{2a}{\kappa^2}\partial_{\xi}{\varphi^{R,\rm e}} + \frac{2}{\kappa^2}\partial^2_{\xi}{\varphi^{R,\rm e}}\right)
e^{-i\omega\tau+i\bm k_\perp\cdot\bm x_\perp}.
\end{align}
We finally have the metric perturbations in the R region in the Regge-Wheeler gauge
\begin{align}
h_{\mu\nu}^R=(\widetilde h_{\mu\nu}^{R,\rm o}+\widetilde h_{\mu\nu}^{R,\rm e})e^{-i\omega\tau+i\bm k_\perp\cdot\bm x_\perp}
\end{align}
with
\begin{align}
\label{hmunuRo}
\widetilde h^{R,\rm o}_{\mu \nu}&=\left(
\begin{array}{cccc}
0
&0&\displaystyle{-\frac{i k_z}{\kappa^2}\partial_{{\xi}}\varphi_{}^{R,\rm o}}&\displaystyle{\frac{i{}k_y}{\kappa^2}\partial_{{\xi}}\varphi_{}^{R,\rm o}}\\
*&0&\displaystyle{-\frac{\omega{}k_z}{\kappa^2}\varphi_{}^{R,\rm o}}&\displaystyle{\frac{\omega{}k_y}{\kappa^2}\varphi_{}^{R,\rm o}}\\
*&*&0&0\\
*&*&*&0
\end{array}\right),
\\
\label{hmunuRe}
\widetilde h^{R,\rm e}_{\mu \nu}&=\left(
\begin{array}{cccc}
\displaystyle{
-\frac{{\cal K}^2(\xi)}{\kappa^2}\varphi_{}^{R,\rm e}-\frac{1}{\kappa^2}\left(\omega^2 \varphi_{}^{R,\rm e}+2a\partial_{{\xi}}\varphi^{R,\rm e}_{}\right)}
&\displaystyle{-\frac{2{}i{}\omega}{\kappa^2}(\partial_{{\xi}}\varphi^{R,\rm e}_{}-a\varphi^{R,\rm e}_{})}&0&0\\
*&\displaystyle{-e^{2a\xi}\varphi_{}^{R,\rm e}-\frac{2}{\kappa^2}\left(a\partial_{{\xi}}\varphi_{}^{R,\rm e}-\partial^2_{{\xi}}\varphi_{}^{R,\rm e}\right)}&0&0\\
*&*&\varphi_{}^{R,\rm e}&0\\
*&*&*&\varphi_{}^{R,\rm e}
\end{array}\right).
\end{align}

%%%%%%%%%%%%%%%%%%%%%%%%%%%%%%%%%%%%%%%%%%%%%%%%%%%%%
\subsection{GWs in left Rindler wedge (L region)}
%%%%%%%%%%%%%%%%%%%%%%%%%%%%%%%%%%%%%%%%%%%%%%%%%%%%%
Similarly, we can obtain the action for the L region.
The line element in the L region is given by
\begin{align}
\label{metricL}
ds^2=e^{2a\widetilde{\xi}}(-d\widetilde{\tau}^2+d\widetilde{\xi}^2)+dy^2+dz^2 \equiv \bar{g}^{L}_{\mu \nu}dx^{\mu}dx^{\nu},
\end{align}
which is related to the Minkowski coordinates through the relation
\begin{align}
\label{coordL}
t=\frac{1}{a}e^{a\widetilde{\xi}}\sinh{a\widetilde{\tau}},\quad
x=-\frac{1}{ a}e^{a\widetilde{\xi}}\cosh{a\widetilde{\tau}},
\end{align}
as shown in Fig.~\ref{fig:epsart}. In a similar way to the case for the 
R region, the metric perturbation is defined by
$g^{L}_{\mu \nu}=\bar{g}^{L}_{\mu \nu}+h^{L}_{\mu \nu}$
with
\begin{align}
h^{L}_{\mu \nu} = 
\begin{pmatrix}
h^{L}_{00}&h^{L}_{01}&\partial_z{\chi}^{L}&-\partial_y{\chi}^{L}\\
h^{L}_{01}&h^{L}_{11}&\partial_z{\psi}^{L}&-\partial_y{\psi}^{L}\\
\partial_z{\chi}^{L}&\partial_z{\psi}^{L}&h^{L}&0\\
-\partial_y{\chi}^{L}&-\partial_y{\psi}^{L}&0&h^{L}
\end{pmatrix}.
\end{align}
We obtain the action for the 
master variable for the odd modes as
\begin{align}
S^{L}_{g(\rm o)} = \frac{1}{32\pi G}\int d\tilde{\tau} d\tilde{\xi} dy dz\left[(\partial_{\tilde{\tau}}\phi^{L})^2 -(\partial_{\tilde{\xi}}\phi^{L})^2 - e^{2a{\tilde{\xi}}}(\partial_\perp \phi^{L})^2\right],
\label{actionSL}
\end{align}
which leads to equation of motion
\begin{align}
\left[-{ \partial_{\tilde{\tau}}^2}+{ \partial_{\tilde{\xi}}^2}
+e^{2a\tilde{\xi}}{ \partial_\perp^2}\right]\phi^L=0.
\label{equationSL}
\end{align}
The action and equation of motion for the even mode master variable $h^L$ are
the same as 
Eqs.~(\ref{actionSL}) and~(\ref{equationSL}), respectively.  
The metric components in the L region are related to $h^{L}$ by
\begin{align}
\partial_\perp^2\psi^{L} = \partial_{\tilde{\eta}}{\phi}^{L},\quad
\partial_\perp^2\chi^{L} = \partial_{\tilde{\xi}}\phi^{L},
\end{align}
and 
\begin{align}
\partial_\perp^2{h^{L}_{00}} &= e^{2a\tilde{\xi}}\partial_\perp^2{h}^{L} + 2a\partial_{\tilde{\xi}}{h^{L}} - 2\partial^2_{\tilde{\tau}}{h}^{L},
\\
\partial_\perp^2{h^{L}_{01}} &=  -2\partial_{\tilde{\tau}}(\partial_{\tilde{\xi}}{h^{L}} - a{h}^{L}),
\\
\partial_\perp^2{h^{L}_{11}} &= -e^{2a\tilde{\xi}}\partial_\perp^2{h}^{L} + 2a\partial_{\tilde{\xi}}{h^{L}} - 2\partial^2_{\tilde{\xi}}{h^{L}}.
\end{align}

Similarly, when $\phi^{L}$ and $h^{L}$ are written in the forms
\begin{align}
\phi^{L}=\varphi^{L,\rm o}(\tilde\xi)e^{-i\omega\tilde\tau-i\bm k_\perp\cdot\bm x_\perp},\quad
h^L=\varphi^{L,\rm e}(\tilde\xi)e^{-i\omega\tilde\tau-i\bm k_\perp\cdot\bm x_\perp}, 
\end{align}
respectively, 
we have the metric perturbations in the L region in the Regge-Wheeler 
gauge
\begin{align}
h_{\mu\nu}^L=(\widetilde h_{\mu\nu}^{L,\rm o}+\widetilde h_{\mu\nu}^{L,\rm e})e^{-i\omega\tilde\tau-i\bm k_\perp\cdot\bm x_\perp}
\end{align}
with
\begin{align}
\label{hmunuLo}
\widetilde h^{L,\rm o}_{\mu \nu}&=\left(
\begin{array}{cccc}
0
&0
&\displaystyle{-\frac{i k_z}{\kappa^2}\partial_{\tilde{\xi}}\varphi_{}^{L,\rm o}}
&\displaystyle{\frac{i k_y}{\kappa^2}\partial_{\tilde{\xi}}\varphi_{}^{L,\rm o}}\\
*&0&\displaystyle{-\frac{\omega k_z}{\kappa^2}\varphi_{}^{L,\rm \rm o}}&\displaystyle{\frac{\omega k_y}{\kappa^2}\varphi_{}^{L,\rm o}}\\
*&*&0&0\\
*&*&*&0
\end{array}\right),
\\
\widetilde h^{L,\rm e}_{\mu \nu}&=\left(
\begin{array}{cccc}
\displaystyle{
-\frac{\tilde{\cal K}^2(\tilde{\xi})}{\kappa^2}\varphi_{}^{L,\rm e}-\frac{1}{\kappa^2}\left(\omega^2 \varphi_{}^{L,\rm e}+2a\partial_{\tilde{\xi}}\varphi^{L,\rm e}_{}\right)}
&\displaystyle{-\frac{2 i \omega}{\kappa^2}(\partial_{\tilde{\xi}}\varphi^{L,\rm e}_{}-a\varphi^{L,\rm e}_{})}
&0
&0\\
*&\displaystyle{-e^{2a\tilde{\xi}}\varphi_{}^{L,\rm e}-\frac{2}{\kappa^2}\left(a\partial_{\tilde{\xi}}\varphi_{}^{L,\rm e}-\partial^2_{\tilde{\xi}}\varphi_{}^{L,\rm e}\right)}&0&0\\
*&*&\varphi^{L,\rm e}_{}&0\\
*&*&*&\varphi^{L,\rm e}_{}
\end{array}\right), 
\label{hmunuLe}
\end{align}
where $\tilde{\cal K}^2(\tilde{\xi})=(\omega^2-\kappa^2e^{2a\tilde{\xi}})$.

\def\phif{\varphi^{F}_{(\lambda)}}
%%%%%%%%%%%%%%%%%%%%%%%%%%%%%%%%%%%%%%%%%%%%%%%%%%
 \section{Quantization of the gravitational waves }
%%%%%%%%%%%%%%%%%%%%%%%%%%%%%%%%%%%%%%%%%%%%%%%%%%
\label{secIII}

Having thus obtained the classical solution for all of the components of the metric perturbations, let us move to consider the quantization of the gravitational waves. As we have shown in the previous section, the master variables are essentially regarded as two decoupled mass- less scalar fields. Therefore, we closely follow the quantization procedure of massless scalar fields living on the Kasner/Rindler metric ~\cite{Entangle}, and the consequences of the quantization of the master variables are very similar to those in the case of a massless scalar field. The results of this section are a review of Ref.~\cite{Entangle}, but it is necessary to explain the new results in Sec. IV.

%%%%%%%%%%%%%%%%%%%%%%%%%%%%%%%%%%%%%%%%%%%%%%%%%%%%%%%
\subsection{Quantization in Kasner spac{e}times (F and P regions)}
%%%%%%%%%%%%%%%%%%%%%%%%%%%%%%%%%%%%%%%%%%%%%%%%%%%%%%%
We perform quantization of the gravitational 
waves described by the actions in the previous section. 
Here we consider the odd and even modes in
the F region (\ref{actionoddF}) and (\ref{actionevenF}). 
For the later convenience, we
introduce the canonically normalized variables
$ \varphi^F_{\rm (o)}=\phi^{F} / \sqrt{16\pi G}$
and $  \varphi^F_{\rm (e)}=h^{F} / \sqrt{16\pi G}$,
so that the corresponding actions reduce to
\begin{align}
S^{F}_{g(\lambda)}= \frac{1}{2}
\int d\eta  d\zeta  dy   dz\left[{({\partial_{\eta}\varphi^{F}_{(\lambda)}})^2} - ({\partial_{\zeta}{\phif}})^2 - e^{2a{\eta}}(\partial_\perp\phif)^2\right],
\end{align}
where $\lambda={\rm o}, {\rm e}$.
The canonical momenta are defined as
\begin{align}
\Pi^{F}_{(\lambda)}(\eta,\zeta,y,z) \equiv \frac{\delta S^{F}_{g(\lambda)}}{\delta \partial_{\eta}{{\phif}}} = 
\partial_{\eta}\varphi^{F}_{(\lambda)}(\eta,\zeta,y,z),
\end{align}
with which the commutation relations are imposed as
\begin{align}
 \label{cor1}
[\hat\varphi^{F}_{(\lambda)}(\eta,\zeta,y,z),\hat{\Pi}^{F}_{(\lambda)}(\eta,\zeta^{\prime},y^{\prime},z^{\prime})] = i\delta(\zeta-\zeta^{\prime})\delta(\vec{x_{\perp}}-\vec{x_{\perp}}^{\prime}),\quad
\text{others}\, =0, 
\end{align}
where $\vec{x_{\perp}} = (y , z) $.

The field operator satisfies the Heisenberg equation of motions
\begin{align}
\left[-\partial^2_{\eta}+\partial^2_{\zeta}
+e^{2a\eta}\partial^2_{\perp}\right]\hat\varphi^{F}_{(\lambda)}=0,
\end{align}
whose
solution can be written as
\begin{align}
\hat\varphi^{F}_{(\lambda)}(\eta,\zeta,y,z) 
=
\int_{-\infty}^{\infty}dk_x\int _{-\infty}^{\infty}d\vec{k_{\perp}} \left[v^{F,\lambda}_{k_x\vec{k}_{\perp}} (\eta,\zeta,y,z)\hat{a}^{F,\lambda}_{k_x\vec{k}_{\perp}} +{\rm h.c.}\right],
\end{align}
where ${\vec{k}_{\perp}} = (k_y , k_z)$, $\hat{a}^{F,\lambda}_{k_x\vec{k}_{\perp}}(\hat{a}^{F,\lambda \dagger}_{k_x\vec{k}_{\perp}}{})$ 
is the annihilation (creation) operators satisfying $\left[\hat{a}_{k_{x} k_{\perp}}^{F, \lambda}, \hat{a}_{k_{k}^{\prime} \boldsymbol{k}_{\perp}^{\prime}}^{F, \lambda \dagger}\right]=\delta\left(k_{x}-k_{x}^{\prime}\right) \delta\left(\boldsymbol{k}_{\perp}-\boldsymbol{k}_{\perp}^{\prime}\right)$ and $v^{F,\lambda}_{k_x,\bm k_\perp}(\eta,\zeta,y,z)$ is
a function written in the form of
 $v^{F,\lambda}_{k_x,\bm k_\perp} =f(\eta) e^{ik_x\zeta}e^{i\vec{k}_{\perp}\cdot\vec{x}_{\perp}}$ with mode functions $f$ satisfying the equation of motion 
\begin{align}
{\partial^2_{\eta} f}(\eta) + k_x^2f(\eta)+e^{2a{\eta}}
|k_\perp|^2 f(\eta) = 0,
\label{modefv}
\end{align}
and the normalization condition
$(\partial_{\eta}f) f^*-f(\partial_{\eta}f^*)=i/(2\pi)^3$.

Let us write
a solution of the equation of motion (\ref{modefv}) % can be chosen as
using the Bessel function of the first kind as~\cite{mode},
\begin{align}
v^{F,\lambda}_{k_x\bm k_\perp}(\eta,\zeta,y,z)= \frac{-i}{2\pi \sqrt{4a\sinh({\pi|k_x|/a})}}  J_{-{i|k_x|}/ {a}} \left(\frac{\kappa e^{a\eta}}{a}\right)
e^{ik_x\zeta} e^{i \vec{k}_{\perp}\cdot\vec{x}_{\perp}}
\equiv\varphi^{F,\lambda}
e^{ik_x\zeta} e^{i \vec{k}_{\perp}\cdot\vec{x}_{\perp}}
\label{anotherB2}
\end{align}
with  $\kappa = |\vec{k_{\perp}}|$. 
This function behaves as $v^{F,\lambda}_{k_x\bm k_\perp}(\eta,\zeta,y,z)\propto e^{-i|k_x|\eta}
e^{ik_x\zeta} e^{i\bm k_\perp\cdot \bm x_\perp}$
in the $\eta\rightarrow -\infty$ limit. 
In Eq.~(\ref{anotherB2}), the latter equality
defines the function $\varphi^{F,\lambda}$ for the 
use in the next section.
This means that this function is a
positive frequency mode function 
near the horizon. 
Furthermore, we note that the mode functions $v^{F,\lambda}_{k_x\bm k_\perp}(\eta,\zeta,y,z)$ with positive $k_x$ represent
right-moving wave modes in the $\zeta$ direction, 
whereas the modes with negative $k_x$ represent left-moving wave modes.
One can see this from the behavior of the mode functions near 
the future horizon~\cite{Entangle}.

One can choose a different solution of Eq.~(\ref{modefv}) 
written in terms of the Hankel function, with which 
the solution 
is written as
\begin{align}
u^{F,\lambda}_{k_x\bm k_\perp}(\eta,\zeta,y,z)= \frac{-i}{
4\pi \sqrt{2a}}  e^{\pi|k_x|/2a} H_{{i|k_x|}/ {a}}^{(2)} \left(\frac{\kappa e^{a\eta}}{a}\right)
e^{ik_x\zeta} e^{i \vec{k}_{\perp}\cdot\vec{x}_{\perp}}.
\end{align}
With this choice, the field operator can be expressed as
\begin{align}
\hat\varphi^{F}_{(\lambda)}(\eta,\zeta,y,z) =
\int_{-\infty}^{\infty}dk_x\int _{-\infty}^{\infty}d\vec{k_{\perp}} \left[u^{F,\lambda}_{k_x\vec{k}_{\perp}} (\eta,\zeta,y,z)
\hat{b}^{F,\lambda}_{k_x\vec{k}_{\perp}} +{\rm h.c.}\right],
\end{align}
where $\hat{b}^{F,\lambda}_{k_x\vec{k}_{\perp}}(\hat{b}^{F,\lambda \dagger}_{k_x\vec{k}_{\perp}}{})$ is the annihilation (creation) operators satisfying
$[\hat{b}^{F,\lambda}_{k_x\vec{k}_{\perp}},{\hat{b}^{{F,\lambda}\dagger}}_{k^{\prime}_x\vec{k}^{\prime}_{\perp}}] =\delta(k_x-k^{\prime}_x)\delta(\vec{k}_{\perp}-\vec{k^{\prime}}_{\perp})$.
The mode functions $u^{F,\lambda}_{k_x\bm k_\perp}(\eta,\zeta,y,z)$
behaves as $u^{F,\lambda}_{k_x\bm k_\perp}(\eta,\zeta,y,z)\propto e^{-i\kappa e^{a\eta}/a}e^{ik_x\zeta} e^{i\bm k_\perp\cdot \bm x_\perp} $ in the
$\eta\to\infty$ limit,  
and therefore this is the natural choice as the 
positive frequency mode at late times. 
As we will show in this section as well as in the appendix \ref{Minv}, $u^{F,\lambda}_{k_x\bm k_\perp}(\eta,\zeta,y,z)$ is the mode function associated with the Minkowski vacuum state. 

Similarly, we can write the same quantized field 
in the P region (see Fig.~\ref{fig:epsart}) in two different ways,
\begin{align}
\hat{\varphi}^{P}_{(\lambda)}(\tilde{\eta},\tilde{\zeta},y,z)  
&= 
\int_{-\infty}^{\infty}dk_x\int _{-\infty}^{\infty}d\vec{k_{\perp}} \left[u^{P,\lambda}_{k_x\bm k_\perp} (\tilde\eta,\tilde\zeta,y,z)
\hat{b}^{P,\lambda}_{k_x\vec{k}_{\perp}} +h.c\right] ,
\\
\hat{\varphi}^{P}_{(\lambda)}(\tilde{\eta},\tilde{\zeta},y,z) &= 
\int_{-\infty}^{\infty}dk_x\int _{-\infty}^{\infty}d\vec{k_{\perp}} \left[v^{P,\lambda}_{k_x\bm k_\perp} (\tilde\eta,\tilde\zeta,y,z)
\hat{a}^{P,\lambda}_{k_x\vec{k}_{\perp}} +{\rm h.c.}\right] ,
\end{align}
where $\hat{a}^{P,\lambda}_{k_x\vec{k}_{\perp}}(\hat{a}^{P,\lambda \dagger}_{k_x\vec{k}_{\perp}}{})$ 
and 
$\hat{b}^{P,\lambda}_{k_x\vec{k}_{\perp}}(\hat{b}^{P,\lambda \dagger}_{k_x\vec{k}_{\perp}}{})$ are the annihilation (creation) 
operators, and the mode functions are
\begin{align}
u^{P,\lambda}_{k_x\bm k_\perp} (\tilde\eta,\tilde\zeta,y,z)&=\frac{i}{4\pi\sqrt{2a}}e^{-\pi |k_x| / 2a} H_{i |k_x| / a}^{(1)}\left(\frac{\kappa e^{-a \tilde{\eta}}}{a}\right)e^{i k_x \tilde{\zeta}}
e^{i \vec{k}_{\perp}\cdot\vec{x}_{\perp}},
\\
v^{P,\lambda}_{k_x\bm k_\perp}(\tilde\eta,\tilde\zeta,y,z)&= \frac{i}{2\pi \sqrt{4a\sinh({\pi|k_x|/a})}}  J_{{i|k_x|}/{a}} \left(\frac{\kappa e^{-a\tilde{\eta}}}{a}\right)
e^{ik_x\tilde{\zeta}}
e^{i \vec{k}_{\perp}\cdot\vec{x}_{\perp}} 
\equiv \varphi^{P,\lambda}
e^{ik_x\tilde{\zeta}}
e^{i \vec{k}_{\perp}\cdot\vec{x}_{\perp}} .
\label{anotherB4}
\end{align}
The modes $v^{P,\lambda}_{k_x\bm k_\perp}(\eta,\zeta,y,z)$ 
with positive $k_x$ represent right-moving wave modes and 
those with negative $k_x$ represent left-moving wave modes
as is discussed in Ref.~\cite{Entangle}.
In Eq.~(\ref{anotherB4}), the latter equality
defines the function $\varphi^{P,\lambda}$ for the convenience in the next section.

As mentioned above, the mode function $u^{F(P),\lambda}_{k_x\bm k_\perp} (\eta,\zeta,y,z)$
that is written in terms of the 
Hankel function in the F(P) region corresponds to the positive frequency mode function 
for the Minkowski vacuum (see Appendix~\ref{Minv}).
This leads us to define the Minkowski vacuum
with the annihilation operator $\hat{b}^{\lambda}_{k_x\vec{k}_{\perp}}$ by
\begin{align}
\hat{b}^{F(K),\lambda}_{k_x\vec{k}_{\perp}} |0\rangle_{M}=0
\end{align}
for any $(k_x,\vec{k}_{\perp})$ and $\lambda$. 
On the other hand,
$v^{F(P),\lambda}_{k_x\bm k_\perp}(\eta,\zeta,y,z)$ is the 
Kasner mode function, with which we can define the Kasner vacuum state by 
\begin{align}
\hat{a}^{F(P),\lambda}_{k_x\vec{k}_{\perp}} |0\rangle_{K}=0
\label{defhatafpl}
\end{align}
for any $(k_x,\vec{k}_{\perp})$ and $\lambda$. 

The Bogoliubov transformation between the two sets of
the mode functions $u^{(F,P)\lambda}_{k_x\bm k_\perp}$ and $v^{(F,P)\lambda}_{k_x\bm k_\perp}$
can be found straightforwardly as follows.
Using some mathematical formula for the Hankel and Bessel functions, 
we arrive at
\begin{align}
\label{modef1}
v_{k_x, \bm k_\perp}^{\mathrm{F}, \lambda}(\eta,\zeta,y,z)&=\frac{-i}{4\pi\sqrt{4a\sinh (\pi |k_x| / a)}}
\left\{e^{\pi |k_x| / a} e^{i k_x \zeta} H_{i |k_x| / a}^{(2)}\left(\frac{\kappa e^{a \eta}}{a}\right)+\left[e^{-i k_x \zeta} H_{i |k_x| / a}^{(2)}\left(\frac{\kappa e^{a \eta}}{a}\right)\right]^{*}\right\}
e^{i \vec{k}_{\perp}\cdot\vec{x}_{\perp}} ,
\\
\label{modef2}
v_{k_x,\bm k_\perp}^{\mathrm{P}, \lambda}(\tilde\eta,\tilde\zeta,y,z)&=\frac{i}{4\pi\sqrt{4a\sinh (\pi |k_x| / a)}} \left\{e^{i k_x \tilde{\zeta}} H_{i |k_x| / a}^{(1)}\left(\frac{\kappa e^{-a\tilde{\eta}}}{a}\right)+e^{-\pi |k_x| / a}\left[e^{-i k_x \tilde{\zeta}} H_{i |k_x| / a}^{(1)}\left(\frac{\kappa e^{-a \tilde{\eta}}}{a}\right)\right]^{*}\right\}
e^{i \vec{k}_{\perp}\cdot\vec{x}_{\perp}} . \nonumber\\
\end{align}
From these equations we can read off the Bogoliubov coefficients as
\begin{align}
v^{F(P),\lambda}_{k_x,\bm k_\perp}(\eta,\zeta,y,z)=\alpha_{k_x} u^{F(P),\lambda}_{k_x, \bm k_\perp}(\eta,\zeta,y,z)+\beta_{-k_x} \left[u^{F(P),\lambda}_{-k_x,-\bm k_\perp}(\eta,\zeta,y,z)\right]^{*},
\end{align}
with
\begin{align}
\alpha_{k_x}=\frac{1}{\sqrt{1-e^{-{2\pi|k_x|}/{a}}}},\quad
\beta_{-k_x} =- \frac {e^{-{{\pi|k_x|}/{a}}}}{\sqrt{1-e^{-{{2\pi|k_x|}/{a}}}}}.
\end{align}
It then follows that
the annihilation and creation operators associated with the 
Minkowski and Kasner modes are related as
\begin{align}
\hat{b}^{F(P),\lambda}_{k_x,\vec{k}_{\perp}}=\left(1-e^{-{{2\pi|k_x|}/{a}}}\right)^{-1/2}\left({\hat{a}^{F(P),\lambda}_{k_x,\vec{k}_{\perp}}-e^{-{{\pi|k_x|}/{a}}}\hat{a}^{F(P),\lambda\dagger}_{-k_x,-\vec{k}_{\perp}}}\right).
\end{align} 

Note that $k_x$ is momentum in the direction of $\zeta$ ($\tilde\zeta$) in the
F-region (P-region), which takes both positive
and negative values. 
The Minkowski vacuum state can be described using the states 
associated with the Kasner vacuum as
\begin{align}
|0\rangle_{M} \propto \exp \left[ \frac{1}{2}\int_{-\infty}^{\infty} d k_x \int_{-\infty}^{{\infty}} d\vec{k}_{\perp} e^{-{\pi|k_x|}/ {a}}a^{F(P),\lambda \dagger}_{k_x,\vec{k}_{\perp}} a^{F(P),\lambda \dagger}_{-k_x,-\vec{k}_{\perp}}\right]|0\rangle_{K}, 
\end{align}
where the Kanser vacuum is defined by Eq.~(\ref{defhatafpl})
for any $(k_x,\vec{k}_{\perp})$ and $\lambda$. 
The above expression can be
rewritten as
\begin{align}
|0\rangle_{M}= \prod_{{k_x}=0}^{\infty}\prod_{\vec{k}_{\perp}=-\infty}^{\infty}\sqrt{1-e^{-{2\pi k_x}/{a}}} \sum_{n=0}^{\infty} e^{-{\pi k_x n}/{a}}|n,\lambda, k_x,\vec{k}_{\perp}\rangle_{K} \otimes|n,\lambda,-k_x,-\vec{k}_{\perp}\rangle_{K},
\label{entangleM}
\end{align}
where $|n,\lambda, k_x,\vec{k}_{\perp}\rangle_{K}=(n !)^{-1 / 2}\bigl(a_{k_x,\vec{k}_{\perp}}^{F(P), \lambda \dagger}\bigr)^{n}|0\rangle_{K}$ is the $n$-th excited state from the Kasner vacuum state $|0\rangle_{K}$
characterized by $k_x,\bm k_\perp$ and $\lambda$.
This expression shows that the Minkowski vacuum is expressed 
as an entangled state between the positive 
and negative momentum modes. 

The expectation value of the number operator
constructed from $\hat{a}^{F(P),\lambda}_{k_x,\vec{k}_{\perp}}$
and $\hat{a}^{F(P),\lambda\dagger}_{k_x,\vec{k}_{\perp}}$
is given by a thermal distribution with the temperature $T=a/2\pi$,
\begin{align}
N_{k_x}\equiv {}_{M}{\langle}0|\hat{a}^{F(P),\lambda\dagger}_{k_x,\vec{k}_{\perp}} \hat{a}^{F(P),\lambda}_{k_x,\vec{k}_{\perp}}|0\rangle_{M}={\frac{1}{{e^{{2\pi |k_x|}/{a}}-1}}}\delta^{(3)}(0),
\end{align}
which is responsible for the entanglement of the Minkowski vacuum state (\ref{entangleM}), where the divergent factor $\delta^{(3)}(0)$ accounts for an infinite spatial volume.
%%%%%%%%%%%%%%%%%%%%%%%%%%%%%%%%%%%%%%%%%%%%%%%%%%%%%%%
\subsection{Quantization in Rindler spacetime (R and L regions)}
%%%%%%%%%%%%%%%%%%%%%%%%%%%%%%%%%%%%%%%%%%%%%%%%%%%%%%%
Quantization of the gravitational waves
in the right and left Rindler wedges
described by the action~(\ref{actionRo}) and~(\ref{actionRe})
can be done in a similar way.
We expand the canonically normalized master variables as
\begin{align}
\hat{{\varphi}}^{R}_{(\lambda)}(\tau,\xi,y,z)&=
\int_{0}^{{\infty}} {d \omega \int_{-\infty}^{{\infty}} d\vec{k}_{\perp}} [v^{R,\lambda}_{\omega,\vec{k}_{\perp}}(\tau,\xi,y,z)\hat{a}^{R,\lambda}_{\omega,\vec{k}_{\perp}}+{\rm h.c.}],
\label{QFR}\\
\hat{{\varphi}}^{L}_{(\lambda)}(\tilde\tau,\tilde\xi,y,z)&=
\int_{0}^{{\infty}} {d \omega \int_{-\infty}^{{\infty}} d\vec{k}_{\perp}} [v^{L,\lambda}_{\omega,\vec{k}_{\perp}}(\tilde\tau,\tilde\xi,y,z)\hat{a}^{L,\lambda}_{\omega,\vec{k}_{\perp}}+{\rm h.c.}],
\end{align}
where we introduced the creation and annihilation operators satisfying $\left[\hat{a}_{\omega k_{\perp}}^{R, \lambda}, \hat{a}_{\omega^{\prime} \boldsymbol{k}_{\perp}^{\prime}}^{R, \lambda \dagger}\right]=\delta\left(\omega-\omega^{\prime}\right) \delta\left(\boldsymbol{k}_{\perp}-\boldsymbol{k}_{\perp}^{\prime}\right)$ and the mode functions are given by
\begin{align}
\label{modeR}
v_{\omega,\bm k_\perp}^{R,\lambda}\left(\tau,\xi,y,z\right)&=\sqrt{\frac{\sinh \pi \omega / a}{4 \pi^{4} a}} e^{-i \omega \tau} K_{i \omega / a}\left(\frac{\kappa e^{a \xi}}{a}\right)e^{i\bm k_\perp\cdot \bm x_\perp}\equiv\varphi^{R,\lambda}
 e^{-i \omega \tau}e^{i\bm k_\perp\cdot \bm x_\perp}
,\\
\label{modeL}
v_{\omega,\bm k_\perp}^{L,\lambda}(\tilde\tau,\tilde\xi,y,z)&=\sqrt{\frac{\sinh \pi \omega / a}{4 \pi^{4} a}} e^{-i \omega \tilde{\tau}} K_{i \omega / a}\left(\frac{\kappa e^{a \tilde{\xi}}}{a}\right)e^{-i\bm k_\perp\cdot \bm x_\perp}\equiv \varphi^{L,\lambda}
 e^{-i \omega \tilde{\tau}}e^{-i\bm k_\perp\cdot \bm x_\perp}.
\end{align}
Here the latter equality defines the function $\varphi^{R,\lambda}$
and $\varphi^{L,\lambda}$ for the use in the next section.
The right (left) Rindler vacuum state is defined by 
\begin{align}
 \hat a^{R(L),\lambda}_{\omega,\bm k_\perp}|0\rangle_{R(L)}=0
\end{align}
for any $\lambda,~\omega,~\bm k_\perp$.

%%%%%%%%%%%%%%%%%%%%%%%%%%%%%%%%%%%%%%%%%%%%%%%%%%%%%%%%%%%%%%
\section{Analytic continuation for metric perturbation components \label{cal}}
%%%%%%%%%%%%%%%%%%%%%%%%%%%%%%%%%%%%%%%%%%%%%%%%%%%%%%%%%%%%%%
\label{analyticcappendix}

In this section, we will
demonstrate that the analytic continuations of the modes from the F and P regions to the R and L regions reproduce the solutions in the latter regions
from those in the former {(and vice versa)}.
The analytic continuations yield identities
between the mode functions and the metric perturbations in the different regions.
The Bogoliubov transformation and the description of the 
Minkowski vacuum in the F and P regions are generalized 
to the entire spacetime including the R and L regions,
leading to the description of the Unruh effect of
metric perturbations. 
Although it has been shown that the analytic continuation works
in the case of a massless scalar field, we should emphasize before proceeding that,
in the case of gravitational waves,
it is not so evident whether or not all the metric components in the different regions
can be connected by means of the analytic continuation.

%%%%%%%%%%%%%%%%%%%%%%%%%%%%%%%%%%%%%%%%%%%%%%%%%%%%%%%%%%%%%%%%%%
\subsection{Analytic continuation of mode functions from Kasner to Rindler}
%%%%%%%%%%%%%%%%%%%%%%%%%%%%%%%%%%%%%%%%%%%%%%%%%%%%%%%%%%%%%%%%%%
We first consider the analytic continuation of the master variables \cite{mode,Entangle}.
The right and left Rindler wedges
are described by the line elements 
(\ref{metricR}) and (\ref{metricL}),
respectively, and their coordinates are related to those of the Minkowski spacetime by 
Eqs.~(\ref{coordR}) and (\ref{coordL})
(see Fig.~\ref{fig:epsart}). The coordinates of the 
four regions, i.e., the F, P, R, and L regions,
are related to each other by the analytic continuations, as summarized in TABLE I.
By inspecting the explicit form of the mode
functions~\eqref{modef1},~\eqref{modef2},~\eqref{modeR}, and~\eqref{modeL},
we find that they are analytically continued as
\begin{eqnarray}
\label{ftor}
v_{-\omega,\bm k_\perp}^{F,\lambda}\left(\eta,\zeta,y,z\right) &{=
\left\{
\begin{array}{cr}v_{\omega,\bm k_\perp}^{R,\lambda}\left(\tau,\xi,y,z\right) 
& ~~ ({\rm R~region})
\\ 
0 & ~~ ({\rm L~region})
\end{array}
\quad\right.} ~ &({\text{for}~ \omega=-k_x>0}),
\\
\label{ftol}
v_{\omega,-\bm k_\perp}^{F,\lambda}\left(\eta,\zeta,y,z\right) &{=
\left\{
\begin{array}{cr}
0
& ~~ ({\rm R~region})
\\ 
v_{\omega,\bm k_\perp}^{L,\lambda}(\tilde\tau,\tilde\xi,y,z) 
 & ~~ ({\rm L~region})
\end{array}
\quad\right.} ~ &({\text{for}~ \omega=k_x>0}),
\\
\label{ptor}
v_{\omega,\bm k_\perp}^{P,\lambda}(\tilde\eta,\tilde\zeta,y,z)&=
{
\left\{
\begin{array}{cr}
  v_{\omega,\bm k_\perp}^{R,\lambda}(\tau,\xi,y,z) 
  & ~\ ({\rm R~region})
\\0 & ~ \ ({\rm R~region})
\end{array}\right.} ~ &({\text{for}~ \omega=k_x>0}),
\\
\label{ptol}
v_{-\omega,-\bm k_\perp}^{P,\lambda}(\tilde\eta,\tilde\zeta,y,z)
&{=
\left\{
\begin{array}{cr}0 & ~\ ({\rm R~region})
\\ v_{\omega,\bm k_\perp}^{L,\lambda}(\tilde\tau,\tilde\xi,y,z) 
& ~ \ ({\rm L~region})
\end{array}\right.} ~ &({\text{for}~ \omega=-k_x>0}),
\end{eqnarray}
where we used the formulas
$K_{-\nu}(z)=K_{\nu}(z), K_{\nu}(z)=-(\pi i / 2) e^{-\nu \pi i / 2} H_{\nu}^{(2)}(e^{-\pi i / 2} z)$, and $K_{\nu}(z)=(\pi i / 2) e^{\nu \pi i / 2} H_{\nu}^{(1)}(e^{\pi i / 2}z)$.
See Ref.~\cite{Entangle} for a more detailed derivation.
Thus, it can be seen that the left-moving (right-moving) modes in the F (P) region
are equivalent to the Rindler modes in the R region, whereas
the right-moving (left-moving) modes in the F (P) region
are equivalent to the Rindler modes in the L region.

We can then identify the creation and annihilation operators in the different
regions as 
\begin{eqnarray}
&&\hat a^{I,\lambda}_{\omega,\vec{k}_{\perp}}\equiv\hat{a}^{F,\lambda}_{-\omega,\vec{k}_{\perp}}=\hat{a}^{P,\lambda}_{\omega,\vec{k}_{\perp}}=\hat{a}^{R,\lambda}_{\omega,\vec{k}_{\perp}}, \label{operatorequivalence1}\\
&&\hat a^{II,\lambda}_{\omega,\vec{k}_{\perp}}\equiv\hat{a}^{F,\lambda}_{\omega,-\vec{k}_{\perp}}=\hat{a}^{P,\lambda}_{-\omega,-\vec{k}_{\perp}}=\hat{a}^{L,\lambda}_{\omega,\vec{k}_{\perp}},
\label{operatorequivalence2}
\end{eqnarray}
which satisfy 
$\bigl[\hat{a}_{\omega, \vec{k}_{\perp}}^{\mathrm{I(II),\lambda}},
\hat{a}_{\omega^{\prime}, {\vec{k}^{\prime}_{\perp}}}^{\mathrm{I(II),\lambda'} \dagger}\bigr]=\delta_{\lambda,\lambda'}\delta (\omega-\omega^{\prime}) \delta (\vec{k}_{\perp}-\vec{k}^{\prime}_{\perp})$ and $
\bigl[\hat{a}_{\omega, \vec{k}_{\perp}}^{\mathrm{I(II),\lambda}}, \hat{a}_{\omega^{\prime},\vec{k}^{\prime}_{\perp}}^{\mathrm{I(II),\lambda'}}\bigr]=\bigl[\hat{a}_{\omega, \vec{k}_{\perp}}^{\mathrm{I(II),\lambda} \dagger}, \hat{a}_{\omega^{\prime}, \vec{k}^{\prime}_{\perp}}^{\mathrm{(I(II),\lambda'} \dagger}\bigr]=0$.

\subsection{Analytic continuation of metric perturbations from F region to R and L regions}

Next, we prove that the metric tensors in the Rindler wedges
are obtained by means of the analytic continuation. A brief summary of the results of this section is presented in Appendix \ref{secv} for the convenience of the readers.

The analytic continuation of the metric tensor consists of the analytic continuation of the mode functions, demonstrated in the previous subsection, and the
transformation of the metric tensor under the coordinate transformation.
This latter step is unique to the case of gravitational waves.
For example, 
under the transformation of the coordinates from the F region to the R region,
the perturbations of the metric tensor transform as
\begin{align}
\label{lowoftransA}
{h}^{F}_{\mu \nu}(x^\prime)=\frac{\partial x^{\rho}}{\partial x^{\prime\mu}}\frac{\partial x^{\sigma}}{\partial x^{\prime\nu }}{h}^{R}_{\rho \sigma}(x),
\end{align}
where $x=\{\tau,\xi, y,z\}$ and $x'=\{\eta,\zeta,y,z\}$.

The odd parity metric perturbations in the F region can be written as
\begin{align}
\tilde{h}^{F,\rm o}_{\mu \nu}(\eta,k_x,\bm k_{\perp})=\frac{1}{\kappa^2}O^{F}_{\mu \nu}(\eta,k_x,\bm k_{\perp})\varphi^{F,\rm o},
\end{align}
where
\begin{align}
O^{F}_{\mu \nu}(\eta,k_x,\bm k_{\perp})
\equiv 
\begin{pmatrix}
0&0&k_x k_z&-k_x k_y\\
*&0&-i k_z \partial_{\eta}&i k_y \partial_{\eta}\\
*&*&0&0\\
*&*&*&0
\end{pmatrix},
\end{align}
and $\varphi^{F,\rm o}$ is defined by 
Eq.~(\ref{anotherB2}) 
or Eq.~(\ref{phiFl}) equivalently.
Similarly, the even parity metric perturbations in the F region can be written as
\begin{align}
\tilde{h}^{F,\rm e}_{\mu \nu}(\eta,k_x,\bm k_{\perp})=\frac{1}{\kappa^2}E^{F}_{\mu \nu}(\eta,k_x,\bm k_{\perp})\varphi^{F,\rm e},
\end{align}
where
\begin{align}
E^{F}_{\mu \nu}(\eta,k_x,\bm k_{\perp})
\equiv 
\begin{pmatrix}
\kappa^2 e^{2a\eta}-2a\partial_{\eta}+2\partial^2_{\eta}&-2i k_xa+2i k_x \partial_{\eta}&0&0\\
*&-(k^2(\eta)+k^2_x)-2a\partial_{\eta}&0&0\\
*&*&\kappa^2&0\\
*&*&*&\kappa^2
\end{pmatrix},
\end{align}
and $\varphi^{F,e}$ is defined by 
Eq.~(\ref{anotherB2}) 
or equivalently Eq.~(\ref{phiFl}).

The metric perturbations in the R region may be expressed in a 
similar way as
\begin{align}
\tilde{h}^{R,\rm o}_{\mu \nu}(\omega,\xi,\bm k_{\perp})=\frac{1}{\kappa^2}O^{R}_{\mu \nu}(\omega,\xi,\bm k_{\perp})\varphi^{R,\rm o}
\end{align}
with
\begin{align}
O^{R}_{\mu \nu}(\omega,\xi,\bm k_{\perp})
\equiv 
\begin{pmatrix}
0&0&-i k_z \partial_{\xi}&i k_y \partial_{\xi}\\
*&0&-\omega k_z&\omega k_y\\
*&*&0&0\\
*&*&*&0
\end{pmatrix}
\end{align}
for the odd modes, and 
\begin{align}
\tilde{h}^{R,\rm e}_{\mu \nu}(\omega,\xi,\bm k_{\perp})=\frac{1}{\kappa^2}E^{R}_{\mu \nu}(\omega,\xi,\bm k_{\perp})\varphi^{R,\rm e},
\end{align}
with
\begin{align}
E^{R}_{\mu \nu}(\omega,\xi,\bm k_{\perp})
\equiv 
\begin{pmatrix}
-({\cal K}^2(\xi)+\omega^2)-2a\partial_{\xi}&2i \omega a-2i \omega \partial_{\xi}&0&0\\
*&-\kappa^2 e^{2a\xi}-2a\partial_{\xi}+2\partial^2_{\xi}&0&0\\
*&*&\kappa^2&0\\
*&*&*&\kappa^2
\end{pmatrix},
\end{align}
for the even modes,
where $\varphi^{R,\rm o}$ and $\varphi^{R,\rm e}$ are defined by 
Eq.~(\ref{modeR}) or Eq.~(\ref{phiRl}).

Now, let us show that $O^F_{\mu\nu}$ and $E^F_{\mu\nu}$ are related,
respectively, to $O^R_{\mu\nu}$ and $E^R_{\mu\nu}$ in the way inferred from
a coordinate transformation.
The coordinate transformation we consider here
is given by the analytic continuation
from the F region to the R region (TABLE I),
\begin{align}
\tau = \zeta -\frac{i\pi}{2a},\quad
\xi = \eta +\frac{i\pi}{2a}\nonumber.
\end{align}
One can directly check that
\begin{align}
O^{F}_{\mu \nu}(\eta,-\omega,\bm k_{\perp})=
\frac{\partial x^\rho}{\partial {x'}^\mu}\frac{\partial x^\sigma}{\partial {x'}^\nu}
O^{R}_{\rho \sigma}(\omega,\xi,\bm k_{\perp}),
\quad 
E^{F}_{\mu \nu}(\eta,-\omega,\bm k_{\perp})=
\frac{\partial x^\rho}{\partial {x'}^\mu}\frac{\partial x^\sigma}{\partial {x'}^\nu}
E^{R}_{\rho \sigma}(\omega,\xi,\bm k_{\perp}).\label{OOEE}
\end{align}
Using this and Eq.~(\ref{ftor}),
we see, for positive $\omega$, that
\begin{align}
&{\widetilde h}_{\mu\nu}^{F,\rm o}(\eta,-\omega,\bm k_\perp)e^{-i\omega\zeta}e^{+i{\bm k}_\perp\cdot {\bm x}_\perp}=\frac{1}{\kappa^2}O^{F}_{\mu \nu}(\eta,-\omega,\bm k_{\perp})v^{F,\rm o}_{-\omega \bm k_\perp}
\nonumber\\
&=
\frac{1}{\kappa^2}
\frac{\partial x^\rho}{\partial {x'}^\mu}\frac{\partial x^\sigma}{\partial {x'}^\nu}
O^{R}_{\rho\sigma}(\omega,\xi,\bm k_{\perp})v^{R,\rm o}_{\omega\bm k_\perp}=
\frac{\partial x^\rho}{\partial {x'}^\mu}\frac{\partial x^\sigma}{\partial {x'}^\nu}
{\widetilde h}_{\rho\sigma}^{R,\rm o}(\omega,\xi,\bm k_\perp)e^{-i\omega\tau}e^{+i{\bm k}_\perp\cdot {\bm x}_\perp},
\\
&{\widetilde h}_{\mu\nu}^{F,\rm e}(\eta,-\omega,\bm k_\perp)e^{-i\omega\zeta}e^{+i{\bm k}_\perp\cdot {\bm x}_\perp}=\frac{1}{\kappa^2}E^{F}_{\mu \nu}(\eta,-\omega,\bm k_{\perp})v^{F,\rm e}_{-\omega \bm k_\perp}
\notag \\ 
& =
\frac{1}{\kappa^2}
\frac{\partial x^\rho}{\partial {x'}^\mu}\frac{\partial x^\sigma}{\partial {x'}^\nu}
E^{R}_{\rho\sigma}(\omega,\xi,\bm k_{\perp})v^{R,\rm e}_{\omega\bm k_\perp}=
\frac{\partial x^\rho}{\partial {x'}^\mu}\frac{\partial x^\sigma}{\partial {x'}^\nu}
{\widetilde h}_{\rho\sigma}^{R,\rm e}(\omega,\xi,\bm k_\perp)e^{-i\omega\tau}e^{+i{\bm k}_\perp\cdot {\bm x}_\perp}.
\end{align}
This result shows that,
by the analytic continuation and the transformation law of the metric
under the coordinate transformation,
all the metric components in the F region can be reproduced from the R region,
and vice versa.

In a similar way, we obtain the metric tensor in the L region by using analytic continuation from the F region.
To this end we write the odd parity metric perturbations
in the L region as
\begin{align}
\tilde{h}^{L,\rm o}_{\mu \nu}(\omega,\tilde{\xi},\bm k_{\perp})=\frac{1}{\kappa^2}O^{L}_{\mu \nu}(\omega,\tilde{\xi},\bm k_{\perp})\varphi^{L,\rm o},
\end{align}
where we defined
\begin{align}
O^{L}_{\mu \nu}(\omega,\tilde{\xi},\bm k_{\perp})
\equiv 
\begin{pmatrix}
0&0&-i k_z \partial_{\tilde{\xi}}&i k_y \partial_{\tilde{\xi}}\\
*&0&-\omega k_z&\omega k_y\\
*&*&0&0\\
*&*&*&0
\end{pmatrix}
\end{align}
and $\varphi^{L,\rm o}$ is defined by 
Eq.~(\ref{modeL}) or Eq.~(\ref{phiLl}).
Similarly, we write the even parity metric perturbations in the L region as
\begin{align}
\tilde{h}^{L,\rm e}_{\mu \nu}(\omega,\tilde{\xi},\bm k_{\perp})=\frac{1}{\kappa^2}E^{L}_{\mu \nu}(\omega,\tilde{\xi},\bm k_{\perp})\varphi^{L,\rm e},
\end{align}
where we defined
\begin{align}
E^{L}_{\mu \nu}(\omega,\tilde{\xi},\bm k_{\perp})
\equiv 
\begin{pmatrix}
-({\cal \tilde{K}}^2(\tilde{\xi})+\omega^2)-2a\partial_{\tilde{\xi}}&2i \omega a-2i \omega \partial_{\tilde{\xi}}&0&0\\
*&-\kappa^2 e^{2a\tilde{\xi}}-2a\partial_{\tilde{\xi}}+2\partial^2_{\tilde{\xi}}&0&0\\
*&*&\kappa^2&0\\
*&*&*&\kappa^2
\end{pmatrix},
\end{align}
and $\varphi^{L,\rm e}$ is defined by Eq.~(\ref{modeL}) or Eq.~(\ref{phiLl}).
The analytic continuation from the F region to the 
L region reads (see TABLE I)
\begin{align}
\tilde{\tau} = -\zeta -\frac{i\pi}{2a},\quad
\tilde{\xi} = \eta +\frac{i\pi}{2a}\nonumber,
\end{align}
using which we can show that the following relations hold:
\begin{align}
O^{F}_{\mu \nu}(\eta,\omega,-\bm k_{\perp})= \frac{\partial x^\rho}{\partial {x'}^\mu}\frac{\partial x^\sigma}{\partial {x'}^\nu}O^{L}_{\rho \sigma}(\omega,\tilde{\xi},\bm k_{\perp}),\quad 
E^{F}_{\mu \nu}(\eta,\omega,-\bm k_{\perp})= \frac{\partial x^\rho}{\partial {x'}^\mu}\frac{\partial x^\sigma}{\partial {x'}^\nu}E^{L}_{\rho \sigma}(\omega,\tilde{\xi},\bm k_{\perp}).
\end{align}
Therefore, by combining the above result and the relation
between the mode functions (\ref{ftol}) 
we obtain, for positive $\omega$,
\begin{align}
&{\widetilde h}_{\mu\nu}^{F,\rm o}(\eta,\omega,-\bm k_\perp)e^{i\omega\zeta}e^{-i{\bm k}_\perp\cdot {\bm x}_\perp}=\frac{1}{\kappa^2}O^{F}_{\mu \nu}(\eta,\omega,-\bm k_{\perp})v^{F,\rm o}_{\omega -\bm k_\perp}\nonumber\\
&=
\frac{1}{\kappa^2}\frac{\partial x^\rho}{\partial {x'}^\mu}\frac{\partial x^\sigma}{\partial {x'}^\nu}O^{L}_{\rho \sigma}(\omega,\tilde{\xi},\bm k_{\perp})v^{L,\rm o}_{\omega\bm k_\perp}=\frac{\partial x^\rho}{\partial {x'}^\mu}\frac{\partial x^\sigma}{\partial {x'}^\nu}{\widetilde h}_{\rho \sigma}^{L,\rm o}(\omega,\tilde{\xi},\bm k_\perp)e^{-i\omega\tilde{\tau}}e^{-i{\bm k}_\perp\cdot {\bm x}_\perp},\\
&{\widetilde h}_{\mu\nu}^{F,\rm e}(\eta,\omega,-\bm k_\perp)e^{i\omega\zeta}e^{-i{\bm k}_\perp\cdot {\bm x}_\perp}=\frac{1}{\kappa^2}E^{F}_{\mu \nu}(\eta,\omega,-\bm k_{\perp})v^{F,\rm e}_{\omega -\bm k_\perp}\nonumber\\
&=
\frac{1}{\kappa^2}\frac{\partial x^\rho}{\partial {x'}^\mu}\frac{\partial x^\sigma}{\partial {x'}^\nu}E^{L}_{\rho \sigma}(\omega,\tilde{\xi},\bm k_{\perp})v^{L,\rm e}_{\omega\bm k_\perp}=\frac{\partial x^\rho}{\partial {x'}^\mu}\frac{\partial x^\sigma}{\partial {x'}^\nu}{\widetilde h}_{\rho\sigma}^{L,\rm e}(\omega,\tilde{\xi},\bm k_\perp)e^{-i\omega\tilde{\tau}}e^{-i{\bm k}_\perp\cdot {\bm x}_\perp},
\end{align}
showing that all the metric components in the L region can be
reproduced from the F region by the analytic continuation and the
transformation law of the metric under the coordinate transformation.

\subsection{Analytic continuation of metric perturbations from P region to R region and L region}
Next let us consider the analytic continuation from the P region to the R region and the L region. 
In the P region, we introduce the following decomposition,
\begin{align}
\tilde{h}^{P,\rm o}_{\mu \nu}(\tilde{\eta},k_x,\bm k_{\perp})=\frac{1}{\kappa^2}O^{P}_{\mu \nu}(\eta,k_x,\bm k_{\perp})\varphi^{P,\rm o},
\end{align}
with
\begin{align}
O^{P}_{\mu \nu}(\tilde{\eta},k_x,\bm k_{\perp}) 
\equiv 
\begin{pmatrix}
0&0&k_x k_z&-k_x k_y\\
*&0&-i k_z \partial_{\tilde{\eta}}&i k_y \partial_{\tilde{\eta}}\\
*&*&0&0\\
*&*&*&0
\end{pmatrix}
\end{align}
and $\varphi^{P,\rm o}$ defined by 
Eq.~(\ref{anotherB4}) or (\ref{phiPl})
for the odd modes, and 
\begin{align}
\tilde{h}^{P,\rm e}_{\mu \nu}(\tilde{\eta},k_x,\bm k_{\perp})=\frac{1}{\kappa^2}E^{P}_{\mu \nu}(\tilde{\eta},k_x,\bm k_{\perp})\varphi^{F,\rm e},
\end{align}
with
\begin{align}
E^{P}_{\mu \nu}(\tilde{\eta},k_x,\bm k_{\perp})
\equiv
\begin{pmatrix}
\kappa^2 e^{-2a\tilde{\eta}}+2a\partial_{\tilde{\eta}}+2\partial^2_{\tilde{\eta}}&2i k_x a+2i k_x \partial_{\tilde{\eta}}&0&0\\
*&-(\tilde{k}^2(\tilde{\eta})+k^2_x)+2a\partial_{\tilde{\eta}}&0&0\\
*&*&\kappa^2&0\\
*&*&*&\kappa^2
\end{pmatrix}
\end{align}
and $\varphi^{P,\rm e}$ defined by Eq.~(\ref{anotherB4}) or Eq.~(\ref{phiPl}) for the even modes. 
In a similar way to the case of the F region, 
using the analytic continuation 
from the P region to the R region,
\begin{align}
\tau = -\tilde{\zeta} -\frac{i\pi}{2a},\quad
\xi = -\tilde{\eta} -\frac{i\pi}{2a}\nonumber,
\end{align}
we obtain the following relations,
\begin{align}
O^{P}_{\mu \nu}(\tilde{\eta},\omega,\bm k_{\perp})= \frac{\partial x^\rho}{\partial {x'}^\mu}\frac{\partial x^\sigma}{\partial {x'}^\nu}O^{R}_{\rho \sigma}(\omega,{\xi},\bm k_{\perp}),\quad 
E^{P}_{\mu \nu}(\tilde{\eta},\omega,\bm k_{\perp})= \frac{\partial x^\rho}{\partial {x'}^\mu}\frac{\partial x^\sigma}{\partial {x'}^\nu}E^{R}_{\rho \sigma}(\omega,{\xi},\bm k_{\perp}),
\end{align}
From this and Eq.~\eqref{ptor},
we obtain, for positive $\omega$,
\begin{align}
&{\widetilde h}_{\mu\nu}^{P,\rm o}(\tilde{\eta},\omega,\bm k_\perp)e^{i\omega\tilde{\zeta}}e^{i{\bm k}_\perp\cdot {\bm x}_\perp}=\frac{1}{\kappa^2}O^{P}_{\mu \nu}(\tilde{\eta},\omega,\bm k_{\perp})v^{P,\rm o}_{\omega \bm k_\perp}\nonumber\\
&=
\frac{1}{\kappa^2}\frac{\partial x^\rho}{\partial {x'}^\mu}\frac{\partial x^\sigma}{\partial {x'}^\nu}O^{R}_{\rho \sigma}(\omega,\tilde{\xi},\bm k_{\perp})v^{R,\rm o}_{\omega\bm k_\perp}=\frac{\partial x^\rho}{\partial {x'}^\mu}\frac{\partial x^\sigma}{\partial {x'}^\nu}{\widetilde h}_{\rho\sigma}^{R,\rm o}(\omega,\tilde{\xi},\bm k_\perp)e^{-i\omega{\tau}}e^{+i{\bm k}_\perp\cdot {\bm x}_\perp},\\
&{\widetilde h}_{\mu\nu}^{P,\rm e}(\tilde{\eta},\omega,\bm k_\perp)e^{i\omega\tilde{\zeta}}e^{i{\bm k}_\perp\cdot {\bm x}_\perp}=\frac{1}{\kappa^2}E^{P}_{\mu \nu}(\tilde{\eta},\omega,\bm k_{\perp})v^{P,\rm e}_{\omega \bm k_\perp}\nonumber\\
&=
\frac{1}{\kappa^2}\frac{\partial x^\rho}{\partial {x'}^\mu}\frac{\partial x^\sigma}{\partial {x'}^\nu}E^{R}_{\rho \sigma}(\omega,{\xi},\bm k_{\perp})v^{R,\rm e}_{\omega\bm k_\perp}=\frac{\partial x^\rho}{\partial {x'}^\mu}\frac{\partial x^\sigma}{\partial {x'}^\nu}{\widetilde h}_{\rho\sigma}^{R,\rm e}(\omega,\tilde{\xi},\bm k_\perp)e^{-i\omega{\tau}}e^{+i{\bm k}_\perp\cdot {\bm x}_\perp}.
\end{align}

In a similar way, we see that the following relations hold with the help of the 
analytic continuation from the P region to the L region,
\begin{align}
O^{P}_{\mu \nu}(\tilde{\eta},-\omega,-\bm k_{\perp})= \frac{\partial x^\rho}{\partial {x'}^\mu}\frac{\partial x^\sigma}{\partial {x'}^\nu}O^{L}_{\rho \sigma}(\omega,\tilde{\xi},\bm k_{\perp}),\quad E^{P}_{\mu \nu}(\tilde{\eta},-\omega,-\bm k_{\perp})= \frac{\partial x^\rho}{\partial {x'}^\mu}\frac{\partial x^\sigma}{\partial {x'}^\nu}E^{L}_{\rho \sigma}(\omega,\tilde{\xi},\bm k_{\perp}),
\end{align}
where the analytic continuation here is given by
\begin{align}
\tilde{\tau} = \tilde{\zeta} -\frac{i\pi}{2a},\quad
\tilde{\xi} = -\tilde{\eta} -\frac{i\pi}{2a}\nonumber.
\end{align}
Then, from the above result and Eq.~(\ref{ptol}) we obtain, for positive $\omega$,
\begin{align}
&{\widetilde h}_{\mu\nu}^{P,\rm o}(\tilde{\eta},-\omega,-\bm k_\perp)e^{-i\omega\tilde{\zeta}}e^{-i{\bm k}_\perp\cdot {\bm x}_\perp}=\frac{1}{\kappa^2}O^{P}_{\mu \nu}(\tilde{\eta},-\omega,-\bm k_{\perp})v^{P,\rm o}_{-\omega -\bm k_\perp}\nonumber\\
&=
\frac{1}{\kappa^2}\frac{\partial x^\rho}{\partial {x'}^\mu}\frac{\partial x^\sigma}{\partial {x'}^\nu}O^{L}_{\rho \sigma}(\omega,\tilde{\xi},\bm k_{\perp})v^{L,\rm o}_{\omega\bm k_\perp}=\frac{\partial x^\rho}{\partial {x'}^\mu}\frac{\partial x^\sigma}{\partial {x'}^\nu}{\widetilde h}_{\rho\sigma}^{L,\rm o}(\omega,\tilde{\xi},\bm k_\perp)e^{-i\omega\tilde{\tau}}e^{-i{\bm k}_\perp\cdot {\bm x}_\perp},\\
&{\widetilde h}_{\mu\nu}^{P,\rm e}(\tilde{\eta},-\omega,-\bm k_\perp)e^{-i\omega\tilde{\zeta}}e^{-i{\bm k}_\perp\cdot {\bm x}_\perp}=\frac{1}{\kappa^2}E^{P}_{\mu \nu}(\tilde{\eta},-\omega,-\bm k_{\perp})v^{P,\rm e}_{-\omega -\bm k_\perp}\nonumber\\
&=
\frac{1}{\kappa^2}\frac{\partial x^\rho}{\partial {x'}^\mu}\frac{\partial x^\sigma}{\partial {x'}^\nu}E^{L}_{\rho \sigma}(\omega,\tilde{\xi},\bm k_{\perp})v^{L,\rm e}_{\omega\bm k_\perp}=\frac{\partial x^\rho}{\partial {x'}^\mu}\frac{\partial x^\sigma}{\partial {x'}^\nu}{\widetilde h}_{\rho\sigma}^{L,\rm e}(\omega,\tilde{\xi},\bm k_\perp)e^{-i\omega\tilde{\tau}}e^{-i{\bm k}_\perp\cdot {\bm x}_\perp}.
\end{align}

Thus, we have shown that the mode functions of the metric perturbations in the R region and the L region are given by the analytic continuations from the F (P) region. 
The results are summarized as follows:
\begin{align}
&{\widetilde h}_{\mu\nu}^{F,\lambda}(\eta,-\omega,\bm k_\perp)e^{-i\omega\zeta}e^{+i{\bm k}_\perp\cdot {\bm x}_\perp}
=
\frac{\partial x^\rho}{\partial {x'}^\mu}\frac{\partial x^\sigma}{\partial {x'}^\nu}{\widetilde h}_{\rho\sigma}^{R,\lambda}(\omega,\xi,\bm k_\perp)e^{-i\omega\tau}e^{+i{\bm k}_\perp\cdot {\bm x}_\perp}
=
{\widetilde h}_{\mu\nu}^{P,\lambda}(\widetilde\eta,\omega,\bm k_\perp)e^{+i\omega\widetilde\zeta}e^{+i{\bm k}_\perp\cdot {\bm x}_\perp},
\label{s1}
\\
&{\widetilde h}_{\mu\nu}^{F,\lambda}(\eta,\omega,-\bm k_\perp)e^{+i\omega\zeta}e^{-i{\bm k}_\perp\cdot {\bm x}_\perp}
=
\frac{\partial x^\rho}{\partial {x'}^\mu}\frac{\partial x^\sigma}{\partial {x'}^\nu}{\widetilde h}_{\rho\sigma}^{L,\lambda}(\omega,\widetilde\xi,\bm k_\perp)e^{-i\omega\widetilde\tau}e^{-i{\bm k}_\perp\cdot {\bm x}_\perp}
=
{\widetilde h}_{\mu\nu}^{P,\lambda}(\widetilde\eta,-\omega,-\bm k_\perp)e^{-i\omega\widetilde\zeta}e^{-i{\bm k}_\perp\cdot {\bm x}_\perp}.
\label{s2}
\end{align}
These results clearly demonstrate the relation
between the modes of gravitational waves in each region:
the left-moving (right-moving) wave modes in the F(P) region
are equivalent to the Rindler modes in the R region, whereas
the right-moving (left-moving) wave modes in the F(P) region
are equivalent to the Rindler modes in the L region.

\subsection{Description of the Minkowski vacuum state and the Unruh effect}
As a result of the analytic continuations from the F (P) region to the R and L regions [Eqs.~(\ref{s1}) and (\ref{s2})], and the equivalence of the operators
[Eqs.~(\ref{operatorequivalence1}) and (\ref{operatorequivalence2})], 
the Minkowski vacuum is described as an entanglement state between the states defined
in the left and right Rindler wedges
and the F and P regions in a unified way, 
\begin{align}
|0\rangle_{M}=\prod_{{\omega}=0}^{\infty}\prod_{\vec{k}_{\perp}=-\infty}^{\infty} \sqrt{1-e^{-{2\pi \omega}/{a}}} \sum_{n=0}^{\infty} e^{-{\pi \omega n}/{a}}|n,\lambda, \omega,\vec{k}_{\perp}\rangle_{I} \otimes|n,\lambda, \omega,\vec{k}_{\perp}\rangle_{II},
\end{align}
where we defined
$|n,\lambda,\omega,\vec{k}_{\perp}\rangle_{I(II)}=(n !)^{-1 / 2}\bigl(a_{\omega,\vec{k}_{\perp}}^{I(II), \lambda \dagger}\bigr)^{n}|0\rangle_{I(II)}$,
and the Rindler vacuum state is defined in each region by
\begin{align}
a_{\omega,\vec{k}_{\perp}}^{I(II), \lambda }|0\rangle_{I(II)}=0  
\end{align}
for any $(\omega,\bm k_\perp)$.
The expectation value of the number operator in the Rindler wedges
reads
\begin{align}
N_{I(II)}\equiv {}_{M}{\langle}0| \hat{a}^{I(II),\lambda\dagger}_{\omega,\vec{k}_{\perp}} \hat{a}^{I(II),\lambda}_{\omega,\vec{k}_{\perp}}|0\rangle_{M}=\frac{1}{{e^{{2\pi \omega}/{a}}-1}}\delta^{(3)}(0),
\end{align}
which shows the thermal distribution with the temperature
determined by the acceleration, $T={a}/{2\pi}$. 
This can be understood as the Unruh effect of the gravitational waves. 

\section{Energy-Momentum Tensor in R region}
\label{secV}
In this section, we
evaluate the energy density of gravitational waves 
 to discuss the difference between the Minkowski vacuum and the Rindler vacuum. 
 We calculate the following effective energy-momentum tensor,
\begin{align}
T_{\mu\nu}^{\rm GW}(x)=-\frac{1}{8\pi G}\langle {}^{(2)}G_{\mu\nu}\rangle,
\end{align}
where $\langle {}^{(2)}G_{\mu\nu}\rangle$ is the second-order part of the Einstein tensor and $\langle \cdots \rangle$ stands for the temporal and spatial average.
Using the expression (\ref{QFR}) in the Rindler wedge, we obtain
\begin{align}
\label{eneden1}
{}_{M}\langle 0|%\rho
\hat{T}^{\rm GW}_{\tau \tau}|0\rangle_{M}&=%\frac{e^{-2a\xi}}{4\pi^2}
\frac{e^{-2a\xi}}{4\pi^2}
\int_{-\infty}^\infty d\omega \frac{\omega^3}{e^{2\pi\omega/a}-1},
\\
\label{eneden2}
{}_{R}\langle0|
\hat{T}^{\rm GW}_{\tau \tau}|0\rangle_{R}&=%\frac{e^{-2a\xi}}{4\pi^2}
\frac{e^{-2a\xi}}{4\pi^2}
\int_{0}^\infty d\omega{\omega^3},
\end{align}
The details of the calculation are presented in Appendix~\ref{eneden}.
Equation~(\ref{eneden1}) shows that the energy density associated with
the Minkowski vacuum state obeys the Planck distribution with the 
temperature $T=a/2\pi$.
The regularized energy momentum tensor is obtained by 
subtracting the expectation values associated with the 
Minkowski vacuum. 
Then, ${}\langle 0_M|
\hat{T}^{GW}_{\tau \tau}|0_M\rangle$ reduces to zero
after regularization, while ${}_{(R)}\langle0|
\hat{T}^{GW}_{\tau \tau}|0\rangle_{(R)}$ reads
\begin{eqnarray}
{}_{R}\langle0|
\hat{T}^{GW}_{\tau \tau}|0\rangle_{R}^{\rm regularized}%_{(R)}{\rm regularized}
&=&
{}_{R}\langle0|
\hat{T}^{\rm GW}_{\tau \tau}|0\rangle_{R}
-{}_{M}\langle 0|%\rho
\hat{T}^{\rm GW}_{\tau \tau}|0\rangle_{M}
\nonumber\\
&=&-\frac{e^{-2a\xi}}{2\pi^2}
\int_{0}^{\infty} d\omega \frac{\omega^3}{e^{2\pi \omega/a}-1}
=-\frac{e^{-2a\xi}}{2\pi^2}\frac{a^4}{240}.
\label{eneden3}
\end{eqnarray}

In the Schwarzschild spacetime
with a mass $M$, whose metric is given by
\begin{align}
ds^2=-(1-{2GM}/{r})dt^2+\frac{dr^2}{(1-{2GM}/{r})}+r^2(d\theta^2+\sin^2\theta d\phi^2)
\end{align}
the expectation value of the energy momentum tensor for a massless scalar field 
with respect to the Boulware vacuum state $|B\rangle$ is given
in the $r\to2GM$ limit by~\cite{Sciama}
\begin{align}
\langle B|\hat{T}^{\nu}_{\mu}|B\rangle_{\rm ren} \sim -\frac{1}{2\pi^2(1-2GM/r)^2}\int_{0}^{\infty}d\omega \frac{\omega^3}{\exp{(2\pi \omega /\kappa)}-1}{\rm diag}\left(-1,\frac{1}{3},\frac{1}{3},\frac{1}{3}\right),
\label{btb}
\end{align}
where $\kappa=1/4GM$. 
This yields an expression for the 
energy density similar to Eq.~(\ref{eneden3}), 
\begin{align}
\langle B|\hat{T}_{00}|B\rangle_{\rm ren} \sim
-\frac{1}{2\pi^2(1-2GM/r)}\int_{0}^{\infty}d\omega \frac{\omega^3}{\exp{(2\pi \omega /\kappa)}-1}=-\frac{1}{2\pi^2(1-2GM/r)}
\frac{\kappa^4}{240}.
\end{align}
One can see a pathological behavior that the renormalized expectation 
value of the energy density diverges at the Schwarzschild
horizon when the Boulware vacuum state is adopted outside the horizon of the Schwarzschild spacetime. 
This is quite similar to the result of Eq.~(\ref{eneden3})
that the energy density diverges at the Rindler horizon when the 
Rindler vacuum state is adopted in the Rindler wedge. 
Thus, the structure of the Rindler vacuum state and the Minkowski vacuum state 
on the Minkowski spacetime is analogous to that
of the Boulware vacuum state and the Hartle-Hawking state on the Schwarzschild spacetime \cite{Sciama}.
The expectation values of the other components of the energy momentum 
tensor will be obtained in a similar way, 
\begin{eqnarray}
{}_{R}\langle0|
\hat{T}^{GW}_{\xi \xi}|0\rangle_{R}^{\rm regularized}
&=&
{}_{R}\langle0|
\hat{T}^{\rm GW}_{\xi \xi}|0\rangle_{R}
-{}_{M}\langle 0|%\rho
\hat{T}^{\rm GW}_{\xi \xi}|0\rangle_{M}
\nonumber\\
&=&-\frac{e^{-2a\xi}}{6\pi^2}
\int_{0}^{\infty} d\omega \frac{\omega^3+4\omega a^2}{e^{2\pi \omega/a}-1}
=-\frac{e^{-2a\xi}}{6\pi^2}\frac{41a^4}{240}
\label{eneden4}\\
{}_{R}\langle0|
\hat{T}^{GW}_{y y}|0\rangle_{R}^{\rm regularized}
&=&{}_{R}\langle0|\hat{T}^{GW}_{z z}|0\rangle_{R}^{\rm regularized}
=
{}_{R}\langle0|
\hat{T}^{\rm GW}_{y y}|0\rangle_{R}
-{}_{M}\langle 0|%\rho
\hat{T}^{\rm GW}_{y y}|0\rangle_{M}
\nonumber\\
&=&-\frac{e^{-4a\xi}}{6\pi^2}
\int_{0}^{\infty} d\omega \frac{\omega^3+\omega a^2}{e^{2\pi \omega/a}-1}
=-\frac{e^{-4a\xi}}{6\pi^2}
\frac{11a^4}{240}
\label{eneden5}
\end{eqnarray}
The result do not satisfy the trace-free property, $\langle
{T}{}^{\mu}{}_\mu\rangle=0$, 
which would be expected by the equation of motion averaged 
within a certain region of spacetime. 
This violation of the trace-free property of the energy momentum tensor could be related to the fact that the gravitational waves violate the
conformal invariance and that the gravitational waves become ambiguous for the modes 
with the wavelength longer than the characteristic scale of the spacetime. The characteristic scale of the Rindler space 
is $1/a$, and the modes of the gravitational waves with the 
wavelength longer than the scale $1/a$ contributes to the property 
$\langle T^\mu{}_\mu\rangle \neq0$. If we take the range of the integration 
with respect to $\omega$ in Eqs.~(\ref{eneden4}) and (\ref{eneden5}) 
to be $\omega$ much larger than $a$, then the trace-free property is 
obtained. 

\section{conclusion}
\label{secVI}
In the present paper, we have derived the solutions of gravitational waves in the future expanding Kasner spacetime (the F region) and the past shrinking Kasner spacetime (the P region) as well as the left (L) and right (R) Rindler wedges in an analytic form. In the derivation, we have performed the metric tensor decomposition in spacetime endowed with two-dimensional plane symmetry and used (an analog of) the Regge-Wheeler gauge. In this formulation, the odd-parity and even-parity modes defined with respect to the parity transformation in the two-dimensional plane are decoupled. We have introduced the two master variables associated with the gravitational-wave degrees of freedom and derived the quadratic actions for them, each of which is equivalent to the action of a massless scalar field in the corresponding background. The master variables were quantized in each region. The mode functions corresponding to the Minkowski vacuum state in the F (P) region were identified and the Bogoliubov transformation between the mode functions of the Minkowski vacuum state and those of the Kasner vacuum state were presented. From the relation it can be seen that the Minkowski vacuum state for the quantized gravitational waves is described as an entangled state constructed on the basis of left-moving and right-moving wave modes associated with the Kasner vacuum state. We have also demonstrated that the metric components of the quantized gravitational waves associated with the Kasner vacuum state in the F and P regions are analytically continued to those in the left and right Rindler wedges. This analytic continuation from the F (P) region to the R (L) region allowed us to see that the Minkowski vacuum state is described as an entangled state between the left and right Rindler states. Our result gives a description of the Unruh effect for gravitational waves in the Rindler wedges. We stress that such an explicit formulation has been done for the first time, and was achieved by extensively using the Regge-Wheeler gauge for a plane- symmetric spacetime.

The description will be useful for investigating the Unruh effect on the vacuum fluctuations of gravitational waves in a frame of uniform acceleration or equivalently in a frame of uniform gravitation by the equivalence principle. A thorough investigation is left for a future study.

%%%%%%%%%%%%%%%%%%%%%%%%%%%%%%%%%%%%
\acknowledgements
We thank T. Inagaki, M. Taniguchi, Y. Matsuo, H. Sakamoto, H. Shimoji, D. Sakuma, K. Ueda, T. Tanaka, A. Higuchi, Y. Nambu, J. Soda,
S. Kanno, A. Matsumura, and H. Suzuki for useful discussions related to the topic 
in the present paper.
This work was partially supported by MEXT/JSPS KAKENHI Grant Numbers~17K05444 (KY),
17H06359 (KY),~JP20H04745 (TK), and~JP20K03936 (TK).

%%%%%%%%%%%%%%%%%%%%%%%%%%%%%%%%%%%%

\begin{appendix}
\section{Positive frequency mode function in the F-region \label{Minv}}
%%%%%%%%%%%%%%%%%%%%%%%%%%%%%%%%%%%%%%%%%%
\noindent
Here we briefly review the positive 
frequency mode function associated with the Minkowski vacuum state 
in the F region \cite{Hankel}.
The relation between the coordinates in the
future expanding Kasner spacetime $(T,\chi)$ and Minkowski spacetime $(t,x)$
is given by
\begin{align}
t=T\cosh \chi,\quad  x=T\sinh \chi.
\end{align}
Here, $T$ and $\chi$ are related to the coordinates used
in the main text, $(\eta,\zeta)$, as
$T=e^{a\eta}/a$ and $\chi=a\zeta$. 
The integral representation for the Hankel function is given by
\begin{align}
e^{-i r\cosh \mathcal{K}+i s \sinh \mathcal{K}}=\frac{1}{2 i} \int_{-\infty}^{\infty} d p e^{-i \mathcal{K} p}\left(\frac{r+s}{r-s}\right)^{i p / 2} e^{\pi p / 2} H_{i p}^{(2)}\left(\left(r^{2}-s^{2}\right)^{1 / 2}\right),
\end{align}
where ${\rm Im}(r \pm s)<0 $.
By transforming the variables as
$s=\kappa x,\ r =\kappa t,\ \mathcal{K}=\sinh^{-1}(q / \kappa)$, we obtain
\begin{align}
e^{-i \omega_{k} t+i q x}=\frac{1}{2 i} \int_{-\infty}^{\infty} d p e^{-i \mathcal{K} p}\left(\frac{t+x}{t-x}\right)^{i p / 2} e^{\pi p / 2} H_{i p}^{(2)}\left(\kappa \left(t^{2}-x^{2}\right)^{1 / 2}\right),
\end{align}
where $\omega_\kappa =\kappa\cosh \mathcal{K}=\sqrt{q^2+\kappa^2}$.
In terms of the coordinates $(T,\chi)$, the right-hand side is written as
\begin{align}
e^{-i \omega_{\kappa} t+i q x}=\frac{1}{2 i} \int_{-\infty}^{\infty} d p e^{-i \mathcal{K} p} e^{i p \chi} e^{\pi p / 2} H_{i p}^{(2)}(\kappa T).
\end{align}
This equation shows that $H^{(2)}_{ip}(\kappa T)$ is indeed
the positive frequency mode function associated with the Minkowski vacuum state.
It is easy to see that
the complex conjugate gives the negative frequency mode function
via the relation
\begin{align}
\left[e^{\pi p/2}H^{(2)}_{ip}(\kappa T)\right]^{*} =e^{-\pi p/2}H^{(1)}_{ip}(\kappa T).
\end{align}

\def\calN{{\sqrt{16\pi G}}}
%%%%%%%%

\section{Quantized GWs: Summary of results}
\label{secv}
%%%%%%%%
We present a summary of the results of the 
quantized gravitational waves in the F, P, R, and L regions.
The quantized metric perturbation in the F region can be written by
introducing $\omega=|k_x|$ as
\begin{align}
&\hat{h}^{F}_{\mu \nu}(\eta,\zeta,y,z)
\nonumber\\
&={\calN}\sum_{\lambda={\rm o,e}}\int_{-\infty}^\infty d k_x \int_{-\infty}^\infty  d^2\vec{k}_{\perp}\left[ {\widetilde{h}}^{F,\lambda}_{\mu \nu}(\eta,k_x,\vec{k}_{\perp})e^{i k_x \zeta+i\vec{k}_{\perp}\cdot \vec{x}_{\perp}}\hat a^{F,\lambda}_{k_x,\bm k_\perp} +{\rm {\rm h.c.}}\right]\nonumber\\
&={\calN}\sum_{\lambda={\rm o,e}}\int_{0}^{\infty}d \omega \int_{-\infty}^\infty d^2\vec{k}_{\perp}\left[{\widetilde{h}}^{F,\lambda}_{\mu \nu}(\eta,-\omega,\vec{k}_{\perp})
e^{-i \omega \zeta+i\vec{k}_{\perp}\cdot \vec{x}_{\perp}}\hat a^{I,\lambda}_{\omega,\bm k_\perp}
+ {\widetilde{h}}^{F,\lambda}_{\mu \nu}(\eta,\omega,-\vec{k}_{\perp})e^{i \omega \zeta-i\vec{k}_{\perp}\cdot \vec{x}_{\perp}}
\hat a^{II,\lambda}_{\omega,\bm k_\perp}
+{\rm {\rm h.c.}}\right],
\end{align}
where the Fourier components $\widetilde h_{\mu\nu}^{F,\rm o}$ and
$\widetilde h_{\mu\nu}^{F,\rm e}$ are given by Eqs.~(\ref{hmunuFo})
and~(\ref{hmunuFe}) with 
\begin{align}
\varphi^{F,\lambda}&= \frac{-i}{2\pi \sqrt{4a\sinh({\pi|k_x|/a})}}  J_{-{i|k_x|}/{a}} \left(\frac{\kappa e^{a\eta}}{a}\right),
\label{phiFl}
\end{align}

We have a similar expression in the P-region
\begin{align}
&\hat{h}^P_{\mu \nu}(\tilde{\eta},\tilde{\zeta},y,z)
\nonumber\\
&={\calN}\sum_{\lambda={\rm o,e}}\int_{-\infty}^\infty d k_x \int_{-\infty}^\infty  d^2\vec{k}_{\perp} \left[{\widetilde{h}}^{P,\lambda}_{\mu \nu}(\tilde{\eta},k_x,\vec{k}_{\perp})e^{i k_x\tilde{\zeta}+i\vec{k}_{\perp}\cdot \vec{x}_{\perp}}a^{P,\lambda}_{k_x,\bm k_\perp} +{\rm {\rm h.c.}}\right]\nonumber\\
&={\calN}\sum_{\lambda={\rm o,e}}\int_{0}^{\infty}d \omega \int_{-\infty}^\infty  d^2\vec{k}_{\perp}\left[{\widetilde{h}}^{P,\lambda}_{\mu \nu}(\tilde{\eta,}\omega,\vec{k}_{\perp})e^{i \omega \tilde{\zeta}+i\vec{k}_{\perp}\cdot \vec{x}_{\perp}}\hat a^{I,\lambda}_{\omega,\bm k_\perp}+{\widetilde{h}}^{P,\lambda}_{\mu \nu}(\tilde{\eta},-\omega,-\vec{k}_{\perp})e^{-i \omega \tilde{\zeta}-i\vec{k}_{\perp}\cdot \vec{x}_{\perp}}\hat a^{II,\lambda}_{\omega,\bm k_\perp}+{\rm {\rm h.c.}}\right].
\end{align}
Here the metric perturbations $\widetilde h^{P,\rm o}_{\mu\nu}$ and $\widetilde h^{P,\rm e}_{\mu\nu}$ are given by Eqs.~(\ref{hmunuPo}) and~(\ref{hmunuPe}), 
respectively, with 
\begin{align}
\varphi^{P,\lambda}= \frac{i}{2\pi \sqrt{4a\sinh({\pi|k_x|/a})}}  J_{{i|k_x|}/{a}}\left(\frac{\kappa e^{-a\tilde{\eta}}}{a}\right),\quad
\label{phiPl}
\end{align}

The quantized tensor mode in the right Rindler wedge is obtained
as
\begin{align}
\hat{h}^{R}_{\mu \nu}(\tau,\xi,y,z)
={\calN}\sum_{\lambda={\rm o,e}}\int_{0}^{\infty}d \omega \int_{-\infty}^\infty d^2\vec{k}_{\perp}\left[{\widetilde{h}}^{R,\lambda}_{\mu \nu}(\omega,\xi,\vec{k}_{\perp})
e^{-i \omega \tau+i\vec{k}_{\perp}\cdot \vec{x}_{\perp}}\hat a^{I,\lambda}_{\omega,\bm k_\perp}
+{\rm {\rm h.c.}}\right],
\end{align}
where the metric perturbations $\widetilde h^{R,\rm o}_{\mu \nu}$ and
$\tilde h^{R,\rm e}_{\mu \nu}$ are given by Eqs.~(\ref{hmunuRo}) and~(\ref{hmunuRe}), respectively, with
\begin{align}
\varphi^{R,\lambda}=\sqrt{\frac{\sinh \pi \omega / a}{4 \pi^{4} a}} e^{-i \omega \tau} K_{i \omega / a}\left(\frac{\kappa e^{a \xi}}{a}\right).
\label{phiRl}
\end{align}

The quantized tensor mode in the left Rindler wedge is obtained
as 
\begin{align}
\hspace{-1cm}\hat{h}^{L}_{\mu \nu}(\widetilde\tau,\tilde\xi,y,z)
={\calN}\sum_{\lambda={\rm o,e}}\int_{0}^{\infty}d \omega \int_{-\infty}^\infty d^2\vec{k}_{\perp}\left[{\widetilde{h}}^{L,\lambda}_{\mu \nu}(\omega,\tilde{\xi},\vec{k}_{\perp})
e^{-i \omega \tilde{\tau}-i\vec{k}_{\perp}\cdot \vec{x}_{\perp}}\hat a^{II,\lambda}_{\omega,\bm k_\perp}
+{\rm {\rm h.c.}}\right],
\end{align}
\def\calk{{\cal K}}
where the metric perturbations $\widetilde h^{L,\rm o}_{\mu \nu}$ and
$\tilde h^{L,\rm e}_{\mu \nu}$ are given by Eqs.~(\ref{hmunuLo}) and~(\ref{hmunuLe}), respectively, with
\begin{align}
\varphi^{L,\lambda}=\sqrt{\frac{\sinh \pi \omega / a}{4 \pi^{4} a}} e^{-i \omega \tilde{\tau}} K_{i \omega / a}\left(\frac{\kappa e^{a \tilde{\xi}}}{a}\right).
\label{phiLl}
\end{align}

%%%%%%%%%%%%%%%%%%%%%%%%%%%%%%%%%%%%%%%%%%%%%%%%%%%%%%%%%%%%%%
\section{Energy density of GWs}\label{eneden}
%%%%%%%%%%%%%%%%%%%%%%%%%%%%%%%%%%%%%%%%%%%%%%%%%%%%%%%%%%%%%%
In this appendix, we derive the vacuum expectation value of the energy density of
GWs [Eqs.~(\ref{eneden1}) and~(\ref{eneden2})].
The energy-momentum tensor of GWs is given by the second-order part of the Einstein tensor as
\begin{align}
T^{\rm GW}_{\mu \nu}=-\frac{1}{8\pi G}\langle {}^{(2)}G_{\mu\nu}\rangle,
\end{align}
where $\langle\cdots\rangle$ denotes spatial and temporal average.
For example, the time-time component of this energy-momentum tensor gives the
energy density of GWs.
Let us now focus on the odd modes in the R region. From a direct computation
with the use of the linear equation of motion~(\ref{EOMR}), we find
\begin{align}
{}^{(2)}G_{\tau \tau}
=-\frac{1}{2}(\partial_{\tau}{\phi}^R)^2+(\text{total derivative}),
\end{align}
where the total derivative will be omitted upon averaging.
We obtain the similar results of other components.
Thus, it can be seen that the energy density can be expressed
simply in terms of the master variable.
It then follows immediately that
\begin{align}
T^{\rm GW}_{\tau \tau}=\frac{1}{16\pi G}
\langle(\partial_{\tau}{\phi}^R)^2\rangle
=\langle (\partial_{\tau}\varphi^R_{\rm (o)})^2\rangle,
\end{align}
where one should recall that $\phi^R =\sqrt{16\pi G}\varphi^R_{\rm (o)}$.

Let us move to the calculation of the vacuum expectation value of the energy density
with respect to the Minkowski vacuum,
\begin{align}
\label{EMT}
{}_{M}\langle 0|\hat{T}^{\rm GW}_{\tau \tau}|0\rangle_{M}={}_{M}\langle 0|(\partial_{\tau}\hat{\varphi}^R_{\rm (o)})^2|0\rangle_{M}.
\end{align}
Substituting Eq.~(\ref{QFR}) into Eq.~(\ref{EMT}), we obtain
\begin{align}
{}_{M}\langle 0|\hat{T}^{\rm GW}_{\tau\tau}|0\rangle_{M}&=
\lim_{x^{\prime}\rightarrow x}\int_{0}^{\infty} d\omega \int d\bm k_{\perp} \left[\frac{\partial_{\tau}v^{R,\rm o}_{\omega,\bm k_{\perp}}(x)\partial_{\tau^{\prime}}v^{R,\rm o*}_{\omega,\bm k_{\perp}}(x^{\prime})}{1-e^{-2\pi \omega/a}}+\frac{\partial_{\tau}v^{R,\rm o*}_{\omega,\bm k_{\perp}}(x)\partial_{\tau^{\prime}}v^{R,\rm o}_{\omega,\bm k_{\perp}}(x^{\prime})}{e^{2\pi \omega/a}-1}\right]\nonumber\\
&=
\lim_{x^{\prime}\rightarrow x}\int_{0}^{\infty} d\omega \omega^2 \int d\bm k_{\perp}\left[\frac{v^{R,\rm o}_{\omega,\bm k_{\perp}}(x)v^{R,\rm o*}_{\omega,\bm k_{\perp}}(x^{\prime})}{1-e^{-2\pi \omega/a}}+\frac{v^{R,\rm o*}_{\omega,\bm k_{\perp}}(x)v^{R,\rm o}_{\omega,\bm k_{\perp}}(x^{\prime})}{e^{2\pi \omega/a}-1}\right],
\end{align}
where we used the following formulas,
\begin{align}
{\label{VEV1}}
{}_{M}\langle0|\hat{a}^{R,\rm o}_{\omega,\bm k_{\perp}}\hat{a}^{R,\rm o\dagger}_{\omega^{\prime},\bm k_{\perp}^{\prime}}|0\rangle_{M}&=\frac{1}{1-e^{-{2\pi \omega}/{a}}}\delta(\omega-\omega^{\prime})\delta(\bm k_{\perp}-\bm k_{\perp}^{\prime}),\\
{\label{VEV2}}
{}_{M}\langle0|\hat{a}^{R,\rm o\dagger}_{\omega,\bm k_{\perp}}\hat{a}^{R,\rm o}_{\omega^{\prime},\bm k_{\perp}^{\prime}}|0\rangle_{M}&=\frac{e^{-{2\pi \omega}/{a}}}{1-e^{-{2\pi \omega}/{a}}}\delta(\omega-\omega^{\prime})\delta(\bm k_{\perp}-\bm k_{\perp}^{\prime}).
\end{align}
Note here that for a technical reason we first consider
two separate spacetime points and then take the coincident limit.
The $\bm{k}_\perp$-integration can be done following Ref.~\cite{Entangle}.
First, we have
\begin{align}
\label{integration}
\int d\bm k_{\perp} v_{\omega,\bm k_{\perp}}^{R,\rm o}(x) v_{\omega,\bm k_{\perp}}^{\mathrm{R,\rm o} *}(x^{\prime})=\frac{\sinh (\pi \omega / a)}{2 \pi^3 a} e^{-i \omega\left(\tau-\tau^{\prime}\right)} \int_{0}^{\infty} d \kappa \kappa K_{i \omega / a}\left(\alpha \kappa \right) K_{i \omega / a}\left(\beta \kappa \right) J_{0}\left(\gamma \kappa \right),
\end{align}
where $\alpha=e^{a\xi}/a,\ \beta=e^{a\xi^{\prime}}/a,\ \gamma=|\bm{x}_{\perp}
-\bm{x}^{\prime}_{\perp}|$,
and we used $K_{-\nu}(z)=K_{\nu}(z)$ and 
$\int_{0}^{2 \pi} d \varphi e^{i \kappa \gamma \cos \varphi}=2 \pi J_{0}\left(\kappa \gamma \right)$.
The integral in the right-hand side can be performed by using
the formula~\cite{formulas, Hankel},
\begin{align}
\int_{0}^{\infty} d \kappa \kappa^{\nu+1} K_{\mu}(\alpha \kappa) K_{\mu}(\beta \kappa) J_{\nu}(\gamma \kappa)=\frac{1}{2} \sqrt{\frac{\pi}{2}} \frac{\gamma^{\nu}}{(\alpha \beta)^{\nu+1}} \Gamma(\nu+\mu+1) \Gamma(\nu-\mu+1)\left(\Theta^{2}-1\right)^{-\nu / 2-1 / 4} \mathcal{B}_{\mu-1 / 2}^{-\nu-1 / 2}(\Theta),
\end{align}
where $ \Theta=(\alpha^{2}+\beta^{2}+\gamma^{2})/2 \alpha \beta$
and $\mathcal{B}_{\mu-1 / 2}^{-\nu-1 / 2}(\zeta)$
(Re$(\mu \pm \nu)>-1$, Re$(\nu)>-1$)
is the associated Legendre function,
\begin{align}
\mathcal{B}_{i \omega / a-1 / 2}^{-1 / 2}(\zeta)=\frac{1}{\sqrt{2 \pi}} \frac{1}{i \omega / a} \frac{1}{\left(\zeta^{2}-1\right)^{1 / 4}}\left[\left(\zeta+\sqrt{\zeta^{2}-1}\right)^{i \omega / a}-\left(\zeta+\sqrt{\zeta^{2}-1}\right)^{-i \omega / a}\right].
\end{align}
Explicitly, in our case we have
$\Theta=(e^{a(\xi-\xi^{\prime})}+e^{-a(\xi-\xi^{\prime})}+a^2{e^{-a(\xi+\xi^{\prime})}}|\bm x_{\perp}-\bm x^{\prime}_{\perp}|^2)/2$ and
\begin{align}
\label{reslutintegral}
&\int_{0}^{\infty} d \kappa \kappa K_{{i \omega}/{a}}\left(\frac{\kappa e^{a \xi}}{a}\right) K_{{i \omega}/{a}}\left(\frac{\kappa e^{a \xi^{\prime}}}{a}\right) J_{0}\left(\gamma \kappa \right)
\nonumber\\
&~~~~=\frac{\pi a^{2} e^{-a (\xi+\xi^{\prime})}}{4 i \sinh (\pi \omega / a)}
\frac{1}{\sqrt{\Theta^{2}-1}}\left[\left(\Theta+\sqrt{\Theta^{2}-1}\right)^{{i \omega}/{a}}-\left(\Theta+\sqrt{\Theta^{2}-1}\right)^{-{i \omega}/{a}}\right].
\end{align}
In the coincidence limit, $x^{\prime}\rightarrow x$ ($\Theta \rightarrow 1$), this reduces to 
\begin{align}
\lim_{x\rightarrow x'}
\int_{0}^{\infty} d \kappa \kappa K_{{i \omega}/{a}}\left(\frac{\kappa e^{a \xi}}{a}\right) K_{{i \omega}/{a}}\left(\frac{\kappa e^{a \xi^{\prime}}}{a}\right) J_{0}\left(\gamma \kappa \right)
=\frac{\pi a\omega e^{-2a\xi}}{2\sinh (\pi \omega / a)},
\end{align}
where we used
\begin{align}
\lim_{\Theta \rightarrow 1}\frac{1}{\sqrt{\Theta^{2}-1}}\left[\left(\Theta+\sqrt{\Theta^{2}-1}\right)^{{i \omega}/{a}}-\left(\Theta+\sqrt{\Theta^{2}-1}\right)^{-{i \omega}/{a}}\right]=\frac{2i\omega}{a}.
\end{align}
Therefore, we finally obtain the following result:
\begin{align}
&
\lim_{x^{\prime}\rightarrow x}\int_{0}^{\infty} d\omega \omega^2 \int d\bm k_{\perp}\left[\frac{v^{R,\rm o}_{\omega,\bm k_{\perp}}(x)v^{R,\rm o*}_{\omega,\bm k_{\perp}}(x^{\prime})}{1-e^{-2\pi \omega/a}}+\frac{v^{R,\rm o*}_{\omega,\bm k_{\perp}}(x)v^{R,\rm o}_{\omega,\bm k_{\perp}}(x^{\prime})}{e^{2\pi \omega/a}-1}\right]
\nonumber\\
&~~~~=
\frac{e^{-2a\xi}}{4\pi^2}
\int_{0}^{\infty} d\omega\left[\frac{\omega^3}{1-e^{-2\pi \omega/a}}+\frac{\omega^3}{e^{2\pi \omega/a}-1}\right]\nonumber\\
&~~~~=
\frac{e^{-2a\xi}}{4\pi^2}
\int_{-\infty}^{\infty} d\omega\left[\frac{\omega^3}{e^{2\pi \omega/a}-1}\right],
\end{align}
where in evaluating the first term we changed
the integration variable as $\omega \rightarrow -\omega$.

In the case of the Rindler vacuum state, it is easy to see that the
vacuum expectation value is given by
\begin{align}
{}_{R}\langle 0|\hat{T}^{\rm GW}_{\tau\tau}|0\rangle_{R}&=
\lim_{x^{\prime}\rightarrow x}\int_{0}^{\infty} d\omega \int d\bm k_{\perp}\partial_{\tau}v^{R,\rm o}_{\omega,\bm k_{\perp}}(x)\partial_{\tau^{\prime}}v^{R,\rm o*}_{\omega,\bm k_{\perp}}(x^{\prime})\nonumber\\
&=
\frac{e^{-2a\xi}}{4\pi^2}
\int_{0}^{\infty} d\omega{\omega^3}.
\end{align}
Then, we obtain the regularized energy density (\ref{eneden3}).
Repeating similar computations, we obtain (\ref{eneden4}) and (\ref{eneden5}), where we use the following results
\begin{align}
\lim_{x^{\prime}\rightarrow x}\frac{\partial}{\partial \xi}\frac{\partial}{\partial \xi^{\prime}}\int d\bm k_{\perp} v_{\omega,\bm k_{\perp}}^{R,\rm o}(x) v_{\omega,\bm k_{\perp}}^{\mathrm{R,\rm o} *}(x^{\prime})&
=\frac{e^{-2a\xi}}{4\pi^2} \frac{(4a^2+\omega^2)}{3}\nonumber,\\
\lim_{x^{\prime}\rightarrow x}\frac{\partial}{\partial y}\frac{\partial}{\partial y^{\prime}}\int d\bm k_{\perp} v_{\omega,\bm k_{\perp}}^{R,\rm o}(x) v_{\omega,\bm k_{\perp}}^{\mathrm{R,\rm o} *}(x^{\prime})&
=\frac{e^{-4a\xi}}{4\pi^2} \frac{(a^2+\omega^2)}{3}\nonumber,\\
\lim_{x^{\prime}\rightarrow x}\frac{\partial}{\partial z}\frac{\partial}{\partial z^{\prime}}\int d\bm k_{\perp} v_{\omega,\bm k_{\perp}}^{R,\rm o}(x) v_{\omega,\bm k_{\perp}}^{\mathrm{R,\rm o} *}(x^{\prime})&
=\frac{e^{-4a\xi}}{4\pi^2} \frac{(a^2+\omega^2)}{3}\nonumber,\\
\lim_{x^{\prime}\rightarrow x}\frac{\partial}{\partial x^{\mu}}\frac{\partial}{\partial x^{\nu \prime}}\int d\bm k_{\perp} v_{\omega,\bm k_{\perp}}^{R,\rm o}(x) v_{\omega,\bm k_{\perp}}^{\mathrm{R,\rm o} *}(x^{\prime})&=0 ~~ ({\text{for $\mu \neq \nu$}}).
\end{align}

\end{appendix}

%%%%%%%%%%%%%%%%%%%%%%%%%%%%%%%%%%%%%%%%%%%%%%%%%%%%%%

\end{document}